\documentclass[acmsmall]{acmart}

\usepackage{tikz}

\usepackage{booktabs}
\makeatletter
\newcommand\footnoteref[1]{\protected@xdef\@thefnmark{\ref{#1}}\@footnotemark}
\makeatother

\usepackage{tabularx}
\usepackage[table]{xcolor}

\AtBeginDocument{%
  }

\setcopyright{acmlicensed}
\copyrightyear{2025}
\acmYear{2025}
\acmDOI{XXXXXXX.XXXXXXX}

\usepackage[textsize=scriptsize]{todonotes}
\setuptodonotes{backgroundcolor=white, bordercolor=white, linecolor=white, textcolor=blue}

\NewDocumentCommand{\inlinenote}{o m}{%
  \fcolorbox{white}{cyan!10}{%
    \parbox{0.9\linewidth}{%
      \setlength{\fboxsep}{1pt}%
      \setlength{\fboxrule}{0.5pt}%
      \IfValueT{#1}{\fcolorbox{cyan}{yellow}{\bfseries\sffamily\scriptsize#1}~}%
      \textcolor{black}{$\blacktriangleright$}~%
      \textit{\textcolor{teal}{#2}}~%
      \textcolor{black}{$\blacktriangleleft$}%
    }%
  }%
}

\begin{document}

\title{Rethinking Services in the Quantum Age: The SOQ Paradigm}


\author{Jose Garcia-Alonso}
\email{jgaralo@unex.es}
\orcid{0000-0002-6819-0299}

\author{Enrique Moguel}
\email{enrique@unex.es}
\orcid{0000-0002-4096-1282}

\author{Jaime Alvarado-Valiente}
\email{jaimeav@unex.es}
\orcid{0000-0003-0140-7788}

\author{Javier Romero-Alvarez} 
\email{jromero@unex.es}
\orcid{0000-0002-3162-1446}

\author{Álvaro M. Aparicio-Morales} 
\email{amapamor@unex.es}
\orcid{0009-0009-5161-5498}

\author{Juan M. Murillo}
\email{juanmamu@unex.es}
\orcid{0000-0003-4961-4030}

\affiliation{%
  \institution{Quercus Software Engineering Group, Universidad de Extremadura}
  \city{Cáceres}
  \country{Spain}
}

\author{Francisco Javier Cavero} 
\email{fcavero@us.es}
\orcid{0009-0004-2453-8814}

\author{Adrián Romero-Flores}
\email{aromero17@us.es}
\orcid{0009-0009-3755-1731}

\author{Alfonso E. Marquez-Chamorro}
\email{amarquez6@us.es}
\orcid{0000-0002-8243-0404}

\author{José Antonio Parejo}
\email{japarejo@us.es}
\orcid{0000-0002-4708-4606}

\author{Antonio Ruiz-Cortés} 
\email{aruiz@us.es}
\orcid{0000-0001-9827-1834}

\affiliation{%
  \institution{I3US Institute, SCORE Lab, Universidad de Sevilla}
  \city{Sevilla}
  \country{Spain}
}

\author{Giuseppe Bisicchia} 
\email{giuseppe.bisicchia@phd.unipi.it}
\orcid{0000-0002-1187-8391}

\author{Alessandro Bocci} 
\email{alessandro.bocci@unipi.it}
\orcid{0000-0002-7000-2103}

\author{Antonio Brogi} 
\email{antonio.brogi@unipi.it}
\orcid{0000-0003-2048-2468}

\affiliation{%
  \institution{University of Pisa}
  \city{Pisa}
  \country{Italy}
}

\renewcommand{\shortauthors}{Jose Garcia-Alonso et al.}

\begin{abstract}
    \textbf{Abstract}. Quantum computing is rapidly progressing from theoretical promise to practical implementation, offering significant computational advantages for tasks in optimization, simulation, cryptography, and machine learning. However, its integration into real-world software systems remains constrained by hardware fragility, platform heterogeneity, and the absence of robust software engineering practices. This paper introduces Service-Oriented Quantum (SOQ), a novel paradigm that reimagines quantum software systems through the lens of classical service-oriented computing. Unlike prior approaches such as Quantum Service-Oriented Computing (QSOC), which treat quantum capabilities as auxiliary components within classical systems, SOQ positions quantum services as autonomous, composable, and interoperable entities. We define the foundational principles of SOQ, propose a layered technology stack to support its realization, and identify the key research and engineering challenges that must be addressed, including interoperability, hybridity, pricing models, service abstractions, and workforce development. This approach is of vital importance for the advancement of quantum technology because it enables the scalable, modular, and interoperable integration of quantum computing into real-world software systems independently and without relying on a dedicated classical environment to manage quantum processing.
\end{abstract}

\ccsdesc[500]{Software and its engineering}
\ccsdesc[500]{Theory of computation Models of computation}

\keywords{Quantum Computing, Quantum Software Engineering, Service-Oriented Quantum}


\maketitle

\newpage
\section{Introduction}
\label{sec:introduction}

Quantum computing, based on the foundational contributions of physicists such as Max Planck \cite{Klein1961} and Niels Bohr \cite{Petersen1963}, has catalyzed groundbreaking innovation across multiple scientific domains by harnessing the fundamental principles of quantum mechanics \cite{Zettili2009}. These principles, such as superposition and entanglement, enable quantum computers to process information in ways that differ radically from classical computational models. As a result, quantum algorithms can achieve exponential speedups for certain classes of problems that are intractable for classical machines \cite{Isaac1998}, leading to the redefinition of computational complexity classes such as Bounded-Error Quantum Polynomial Time (BQP) \cite{Aaronson2010}. These emerging capabilities are already beginning to reshape practical fields, including finance \cite{Orus2019}, pharmacogenetics \cite{Romero2021}, combinatorial optimization \cite{Heng2022}, and Artificial Intelligence \cite{Sood2024}.

Despite the potential, quantum computing remains in its early stages, particularly with the current generation of Noisy Intermediate-Scale Quantum (NISQ) devices \cite{Zhao2020}. These systems are error-prone, offer limited qubit counts, and require hybrid execution models where quantum and classical computing are tightly linked \cite{Leymann2020}. To access these quantum capabilities, researchers and developers primarily use cloud platforms offered by companies such as IBM, Google, Amazon, and Microsoft \cite{Zhao2020}. These platforms enable experimentation and application development through on-demand access to quantum processors, simulators, and hybrid computing environments, democratizing quantum computing for academia and industry alike \cite{Murillo2025}.

However, to take full advantage of quantum computing in real-world systems, it is necessary to integrate it with existing software architectures and paradigms. Service-Oriented Computing (SOC), which has long been a foundation for building scalable, modular, and interoperable software systems, provides a promising path for this integration \cite{Moguel2022}. The first efforts in this direction defined Quantum Service-Oriented Computing (QSOC) \cite{Kumara2021,Moguel2022}, which adapted SOC principles to expose quantum capabilities as abstract and composable services. This adaptation aimed to abstract the technical complexity of quantum backends and integrate them as auxiliary components within classical service architectures \cite{Romero2024b}. QSOC played a foundational role in fostering interoperability, modularity, and reuse in quantum applications, particularly in hybrid computing environments.

Additionally, quantum computing is now maturing to the point where it no longer needs to be framed as an extension to classical systems. The time has come to treat quantum computing not just as a compatible addition to SOC, but as a first-class citizen within service-oriented ecosystems. To this end, we introduce the concept of Service-Oriented Quantum (SOQ), a paradigm that directly inherits the principles of SOC, such as encapsulation, loose coupling, interoperability, and dynamic composition, and applies them to quantum computing in a native way.

\begin{figure}[!ht]
\centering
    \caption{Transition to Service-Oriented Quantum (SOQ)}
    \includegraphics[width=0.95\textwidth]{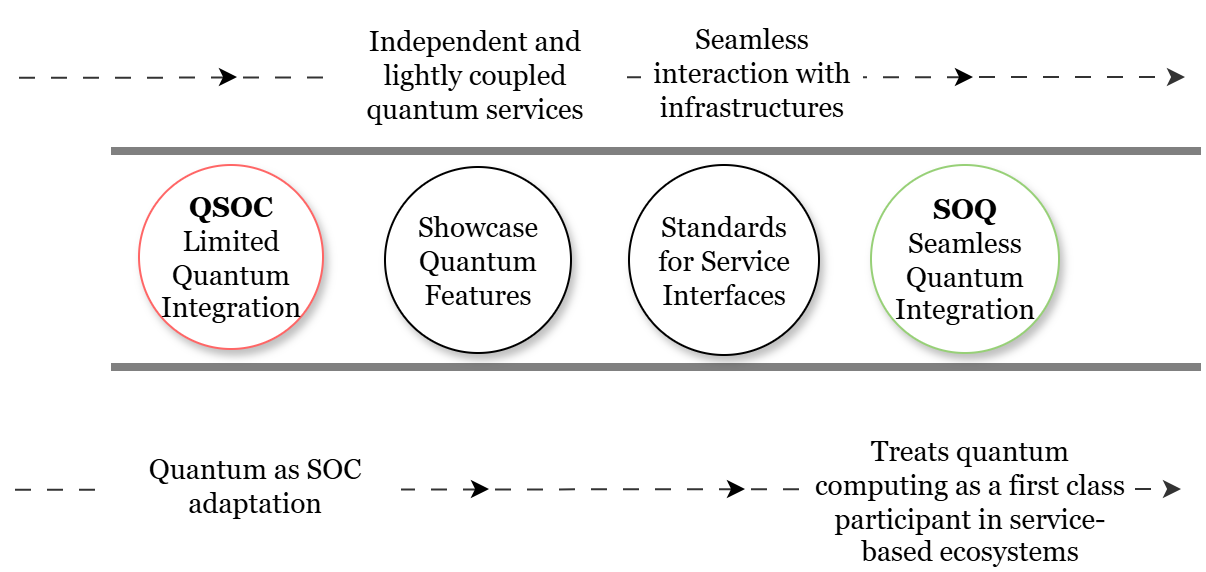}
    \label{fig:QSOC_SOQ}
\end{figure}

SOQ envisions a world where quantum capabilities are exposed, discovered, composed, and orchestrated just like any other service, without requiring them to be anchored in classical service ecosystems. In contrast to QSOC, which emphasized adapting classical methods to incorporate the quantum paradigm, SOQ proposes a dual-native model where classical and quantum services are peers in the architecture. SOQ services are inherently modular and platform-independent, enabling hybrid applications to dynamically interact with quantum and classical backends, optimizing for cost, fidelity, and performance. In short, SOQ inherits and directly applies traditional service-oriented principles, treating quantum computing as a first-class participant in service-based ecosystems, as seen in the transition reflected in Fig. \ref{fig:QSOC_SOQ}.

This work aims to lay the theoretical and architectural foundations of SOQ, analyze its differences with respect to previous models such as QSOC, and identify the key challenges and opportunities it brings. Addresses interoperability across quantum platforms, dynamic pricing, hybrid orchestration, and workforce development, critical pillars for enabling SOQ at scale. From this perspective, SOQ is positioned not only as a transitional concept for current NISQ-era systems but as a durable model that will persist into the Fault-Tolerant Quantum Computing (FTQC) era.

The structure of the paper is as follows. Section \ref{sec:fundamentals} introduces the foundational concepts and technological components that define SOQ, including quantum computing, software engineering principles, and the hybrid execution model. Section \ref{sec:interest} outlines a set of research and engineering challenges that must be addressed to fully realize SOQ, spanning interoperability, platform independence, pricing, and workforce training. Section \ref{sec:influence} presents concrete case studies and real-world scenarios that demonstrate the relevance of these challenges across diverse application domains. Section \ref{sec:challenges} reviews existing work on quantum service abstraction, and we present the challenges that must be overcome to achieve SOQ. Section \ref{sec:soq} contrasts SOQ with the earlier paradigm of Quantum Service-Oriented Computing (QSOC), highlighting the conceptual and architectural differences, and presenting a proposed technology stack for SOQ. Section \ref{sec:conclusion} concludes the paper with a synthesis of insights and directions for future work.

\section{QSOC Fundamentals}
\label{sec:fundamentals}




To lay a solid foundation for SOQ, we must revisit and clarify the key concepts that distinguish this approach from traditional QSOC. Rather than viewing quantum services merely as extensions of classical Service-Oriented Computing, SOQ posits quantum services as native, independent, and composable building blocks in service-oriented architectures. This section defines the core principles underpinning SOQ, focusing on quantum computing, quantum software engineering, hybridization, services, platforms, and economic models.

\subsection{Quantum Computing}

Quantum computing is an emerging computational paradigm based on the principles of quantum mechanics, which allow information to be processed using quantum bits (qubits) that exist in superposition and can become entangled. These characteristics enable quantum computers to perform certain types of calculations exponentially faster than classical computers.

The foundational concepts of quantum computation were established in the 1980s by pioneers such as Richard Feynman, who proposed using quantum systems to simulate physical phenomena that classical computers cannot model efficiently \cite{Feynman1982}, and David Deutsch, who introduced the notion of a universal quantum computer \cite{Deutsch1985}. Quantum algorithms such as Shor’s algorithm for integer factorization \cite{Shor1997} and Grover’s algorithm for unstructured search \cite{Grover1996} provided the first concrete demonstrations of quantum speedup.

Quantum computing operates on qubits, which can represent a combination of both 0 and 1 states due to superposition. When multiple qubits become entangled, the state of one qubit becomes dependent on the state of another, even at a distance, a property that is crucial for parallelism and interference in quantum computation \cite{Nielsen2010}. These mechanisms allow quantum computers to evaluate multiple possibilities simultaneously and eliminate incorrect outcomes through quantum interference.

However, practical quantum computing faces significant challenges \cite{Murillo2025}. Current devices, known as NISQ computers \cite{Preskill2018}, are susceptible to noise, decoherence, and gate errors, which limit the depth and reliability of quantum circuits. As such, quantum software must be designed with an awareness of the hardware's physical limitations and the probabilistic nature of measurement.

\subsection{Quantum Software Engineering}

Quantum Software Engineering (QSE) is an emerging field at the intersection of software engineering and quantum computing \cite{Murillo2025}, focused on the systematic development, testing, deployment, and maintenance of software systems that involve quantum components. Unlike classical software, quantum software must account for the non-deterministic, probabilistic, and resource-constrained nature of quantum hardware.

Early mentions of QSE framed it as a grand challenge for the discipline \cite{Clark2002}, anticipating that the rise of practical quantum computers would necessitate entirely new software engineering practices. More recently, researchers have begun to define the scope and processes of QSE, identifying the need for quantum-specific methods in requirements engineering, architecture, programming languages, verification, and maintenance \cite{Zhao2020,Piattini2020}.

One of the main distinguishing features of QSE is the inherent hybridity of most real-world quantum software. Current quantum algorithms (e.g., Variational Quantum Eigensolver (VQE), Quantum Approximate Optimization Algorithm (QAOA)) are executed partially on classical processors that coordinate and optimize quantum workloads. As a result, actually quantum software is not isolated, it must be tightly integrated with classical components, often requiring new software architectures and orchestration models to coordinate quantum and classical execution.

On the other hand, quantum programs are executed on hardware that returns probabilistic results, and measurements collapse the state of qubits, making it impossible to inspect intermediate quantum states without altering them \cite{Knill2010}. This limits the applicability of traditional debugging techniques and motivates the development of new validation methods based on statistical testing, simulations, and formal verification.

Furthermore, the lack of mature development environments, IDEs, and toolchains hinders productivity \cite{Zhao2020}. While frameworks such as Qiskit, Cirq, and Q\# provide quantum programming capabilities, they lack many of the lifecycle management tools expected in classical software development \cite{Weder2022}. The need for versioning, CI/CD pipelines, modular design practices, and service-based reuse in quantum development is widely acknowledged by the QSE community.

\subsection{Hybrid Classical-Quantum Systems}

In the current landscape of quantum computing, hybrid classical-quantum systems represent the dominant execution model \cite{Moguel2022,valencia2021hybrid}. Due to the limited capabilities of present-day NISQ devices, characterized by short coherence times, gate errors, and restricted qubit counts, quantum computers are not yet capable of solving most problems independently. Instead, they operate in tandem with classical processors that manage the control flow, data preprocessing, optimization loops, and result interpretation.

A hybrid classical-quantum system typically follows a workflow partitioning model, where specific computational tasks are assigned to quantum processors, such as matrix exponentiation, state preparation, or sampling, while the broader orchestration, including control structures and decision-making logic, remains in the classical domain \cite{Bharti2022}. This approach enables developers to benefit from quantum acceleration in critical subroutines while retaining the reliability and scalability of classical computing for the rest of the application.

Well-known examples of this pattern include VQAs such as the VQE and the QAOA. These algorithms execute parameterized quantum circuits whose outputs are evaluated by a classical optimizer that adjusts the quantum parameters iteratively \cite{McClean2016}. This feedback loop exemplifies the tight interdependence between classical and quantum components and illustrates the need for hybrid-aware software architectures.

From a software engineering perspective, the hybrid model introduces several key challenges. First, it is essential to preserve modular separation of concerns, even when classical control logic and quantum operations are tightly interdependent. Second, orchestration layers must accommodate heterogeneous execution paradigms, managing synchronous classical function calls alongside asynchronous quantum job submissions, while ensuring robust scheduling, retry, and failure-handling mechanisms. Finally, data interoperability becomes critical, particularly in the transformation of classical data into quantum states, such as through amplitude or angle encoding, and the decoding of quantum measurement outputs into usable classical formats for downstream processing and decision-making.

\subsection{Service}

In the field of Software Engineering, a service is a self-contained, platform-agnostic computational unit that exposes a set of functionalities through well-defined interfaces \cite{Wei2010}. This concept lies at the heart of SOC, which promotes the decomposition of software systems into modular, reusable, and loosely coupled services that can be dynamically composed and orchestrated \cite{Moguel2022}. Services are typically designed to be stateless, autonomous, and discoverable, enabling systems to scale, evolve, and integrate flexibly across heterogeneous platforms.

In modern architectures, particularly in cloud-native and microservice-based systems, services may encapsulate business logic, data access, external APIs, or computational models. They are published through service registries, invoked through standard protocols (e.g., REST, gRPC, SOAP), and governed by Service-Level Agreements (SLAs) that define guarantees regarding performance, availability, and cost \cite{GamezDiaz2018}.

From an engineering perspective, services provide:

\begin{itemize}
    \item Encapsulation. Implementation details are hidden behind standardized interfaces.

    \item Interoperability. Services can interact across platforms and languages using agreed-upon protocols and data formats.

    \item Composability. Services can be orchestrated into higher-level workflows or composite applications.

    \item Reusability. The same service can be used across different domains or projects with minimal adaptation.
\end{itemize}

\subsection{Quantum Service}

A quantum service is a modular, remotely accessible unit of functionality that encapsulates quantum capabilities and exposes them through standardized interfaces. Just as traditional services in SOC abstract complex business or computational logic, quantum services abstract quantum computation, enabling users to interact with quantum systems without needing to manage the underlying quantum hardware, circuit representations, or error correction mechanisms.

Quantum services are central to the QSOC paradigm. They allow quantum operations, such as circuit execution, quantum simulations, entanglement generation, or quantum-enhanced optimization, to be integrated into distributed applications as invocable services, decoupled from specific hardware backends or programming languages. This abstraction is essential for building interoperable, platform-independent, and reusable quantum software components.

A typical quantum service has the following characteristics:

\begin{itemize}
    \item Standardized interface (API). The service provides a contract that defines accepted inputs (e.g., circuits, parameters, quantum jobs) and expected outputs (e.g., measurement results, fidelities, state vectors). These interfaces are commonly implemented using RESTful APIs, OpenAPI specifications, or SDKs in languages like Python or Q\# \cite{Romero2023,romero2022using}.

    \item Abstracted execution model. The service may execute on real quantum hardware (QPUs), simulators, or emulators. Users do not need to know the physical platform used unless explicitly required. The abstraction of the execution model is not complete, and services often need to expose those details, whereas in SOQ this could be avoided by enhancing the seamless integration between quantum software and its invocation \cite{Romero2023}.

    \item Metadata and constraints. The service typically includes metadata such as the qubit count, noise level, execution time, number of shots, or queue status. These non-functional properties are essential for orchestrating quantum services within larger workflows.

    \item Security and isolation, like any remote service \cite{Zhang2014}, quantum services must support authentication, authorization, and isolation of workloads. These concerns become critical when multiple users or applications share access to limited QPU resources.
\end{itemize}

A prominent example of real-world quantum service is Amazon Braket, which allows users to define quantum tasks (several shots of a circuit to be executed in a simulator or quantum computer) in a provider-neutral format and run them on hardware from different vendors. Similarly, IBM Quantum Services exposes quantum circuits through Qiskit APIs, enabling cloud-based quantum computation and integration with classical pipelines. These platforms follow the service model by abstracting back-end complexity and offering programmable quantum capabilities through accessible interfaces \cite{AlvaradoValiente2024}.

From a Software Engineering viewpoint, quantum services introduce both new requirements and unique challenges \cite{Murillo2025,Leite2025,aparicio2024overview}. Input and output abstractions must accommodate quantum-specific constructs such as superposition and entanglement, while maintaining compatibility with classical data formats to ensure seamless integration. Additionally, error mitigation strategies may need to be embedded within the service layer itself, enabling users to define fidelity targets or trade-offs as part of service invocation. Furthermore, quantum services must expose cost and resource constraints, such as qubit usage, execution time, and queue availability, through the service interface or service-level agreements (SLAs), allowing for informed decision-making and negotiation in distributed quantum applications.

\subsection{Platform}

In the context of QSOC, a platform refers to the underlying technological environment that enables the execution, composition, and management of quantum services \cite{Nguyen2024cloud}. While in classical service-oriented computing a platform may consist of cloud infrastructure, runtime environments, and service orchestration frameworks, quantum platforms add several layers of complexity due to the heterogeneous nature of quantum hardware, the immaturity of tooling, and the physical constraints of quantum processors.

A typical quantum computing platform includes:

\begin{itemize}
    \item Quantum hardware (QPUs). Devices built on various technologies such as superconducting qubits (e.g., IBM, Rigetti), trapped ions (e.g., IonQ, Honeywell), photonic qubits (e.g., Xanadu), or neutral atoms. Each type offers distinct trade-offs in terms of coherence time, gate fidelity, connectivity, and qubit scalability \cite{Ladd2010}.

    \item Execution backends. In addition to real QPUs, platforms often offer high-fidelity simulators or emulators, which are essential for testing and benchmarking quantum algorithms under ideal or noisy conditions. These backends can typically be invoked through the same API interfaces as real devices, allowing seamless migration between testing and production environments.

    \item Software stacks and SDKs. Platforms provide development environments (e.g., Qiskit, Cirq, Q\#) that include tools for circuit design, compilation, transpilation, optimization, and visualization. These tools are essential for preparing circuits in forms compatible with target hardware constraints (e.g., limited gate sets, qubit topology) \cite{Faro2023,Serrano2022}.

    \item Middleware and orchestration. This includes job schedulers, resource managers, and queueing systems that manage user submissions, ensure fairness, and optimize execution across users and workloads. In multi-tenant cloud environments, these middleware layers enforce quality of service (QoS) policies and enable metering for cost-based access models \cite{Faro2023}.

    \item Monitoring and metering. Platforms offer dashboards and APIs for tracking usage metrics, queue positions, job outcomes, error rates, and hardware availability. These features support both performance engineering and cost management in production environments.
\end{itemize}

\subsection{Pricing}

In classical SOC, pricing models are a well-established mechanism that governs the economic interaction between service providers and consumers \cite{GamezDiaz2018,GarciaFernandez2024}. These models are typically tied to cloud-based infrastructures, where computational resources, such as CPU time, memory, storage, and network bandwidth, are offered as services on a pay-per-use basis. Common pricing strategies include subscription models, tiered pricing, on-demand billing, and reserved capacity, all of which aim to provide flexibility and predictability for users while optimizing infrastructure utilization for providers.

The design of pricing models in classical computing takes into account several factors, such as:

\begin{itemize}
    \item Resource consumption: execution time, memory, I/O operations, bandwidth, among others.

    \item Quality of service (QoS): availability, latency, and fault tolerance.

    \item Service priority: faster or guaranteed execution for premium users.

    \item Scalability: ability to provision and de-provision resources dynamically based on demand.
\end{itemize}

This pricing logic is deeply embedded in service orchestration and cloud-native architectures, where it influences deployment strategies or load balancing \cite{alvarado2024orchestration}.  Beyond infrastructure-level resources, SOC also encompasses higher-level services, particularly Software as a Service (SaaS), where applications are delivered either through user interfaces (e.g., web front-ends) or programmatic access (APIs). Pricing in this context is often determined by features plans and add-ons \cite{GarciaFernandez2024}, and operational limits such as request volume, and number of active users \cite{GamezDiaz2018}. These models extend the principles of flexibility and predictability beyond infrastructure, shaping the economic dimension of application-level services \cite{fresno2022}.

In the emerging field of QSOC, pricing introduces a new layer of complexity. Quantum computing resources, unlike classical infrastructure, are scarce, expensive, and physically constrained. Quantum processors require specialized environments (e.g., cryogenics), have limited qubit counts, and are prone to errors and decoherence, making them significantly more costly to operate and maintain. As a result, quantum services made available through QSOC must adopt customized pricing schemes that reflect these unique operational realities \cite{RuizCortes2025}.

Current quantum providers (e.g., IBM, Amazon Braket, IonQ) typically charge based on:

\begin{itemize}
    \item Number of shots (i.e., repeated circuit executions) and/or number of tasks.

    \item Qubit usage and circuit depth.

    \item Number of 1- and 2-qubit gate operations.

    \item Access tier (e.g., free tier, standard, priority queue).

    \item Backend type (simulator vs. real quantum hardware).
\end{itemize}

In QSOC, where quantum functionalities are exposed as services within classical architectures, pricing must also address the integration of quantum and classical components. This may include billing for hybrid workflows that span multiple execution environments or combining quantum processing with classical preprocessing, orchestration, and result handling. Furthermore, given the non-deterministic nature of quantum outputs and the need for repeated executions, pricing may also incorporate probabilistic guarantees or statistical convergence criteria.

Ultimately, in both classical SOC and QSOC, pricing is not just an economic mechanism, but also a driver of resource optimization, service selection, and execution planning. As quantum services become more prevalent within service-oriented ecosystems, pricing strategies will need to evolve to balance fair access, performance guarantees, and operational cost recovery.


\section{Interest in Quantum Service-Oriented Computing}
\label{sec:interest}

Increased efforts have focused on integrating quantum computing with service-oriented principles \cite{Romero2024b} to address challenges such as interoperability, demand management, and platform independence. This is a direct consequence of the rapid advancement of quantum computing and the growing need for structured solutions. The term SOQ is a novel proposition defined in this work, distinct from the existing QSOC concept. While some early references to QSOC exist from the mid-2010s, due to the 2016 milestone of IBM providing the first quantum computers accessible via the cloud, a more notable mention of QSOC within the field of Quantum Software Engineering appeared in the early 2020s, coinciding with the increasing number of cloud-based quantum computing platforms available.

This has led to a growing interest in this field, with a rise in academic conferences and workshops highlighting the pivotal role of the development of QSOC. The International Conference on Service-Oriented Computing (ICSOC)\footnote{\url{https://icsoc2025.hit.edu.cn/}} serves as a prime example of including \textit{Quantum Service Computing} as one of its topics of interest, which reflects the expanding academic activity in the QSOC research area within the service computing community. The convergence of quantum computing and service-oriented architectures has been a key research topic at several other events. This includes the IEEE International Conference on Quantum Software (IEEE QSW)\footnote{\url{https://services.conferences.computer.org/2025/qsw}}; the International Conference on Software Engineering (ICSE)\footnote{\url{https://conf.researchr.org/home/icse-2025}}; and workshops such as the International Workshop on Quantum Software Engineering and Technology (Q-SET)\footnote{\url{https://qserv.spilab.es/q-set-2025-home}}, which is part of the IEEE Quantum Week. Furthermore, events such as the Symposium and Summer School On Service-Oriented Computing (SummerSOC)\footnote{\url{https://www.summersoc.eu}} also underscore the escalating interest in this field, among many others.

Moreover, QSOC-related research topics have been included in some principal journals because of this increasing preoccupation. For instance, the IEEE Transactions on Services Computing and ACM Transactions on Software Engineering and Methodology are two examples where articles exploring the architectural and implementation challenges of quantum services have been published. This trend is also evident in several journals that have dedicated special issues to topics in quantum software, including QSOC, such as the Journal of Systems and Software, the Journal of Information and Software Technology and the Journal Quantum Information Processing, among many others. 

\begin{figure}[!ht]
\centering
\caption{QSOC publications over time.}
    \includegraphics[width=0.9\textwidth]{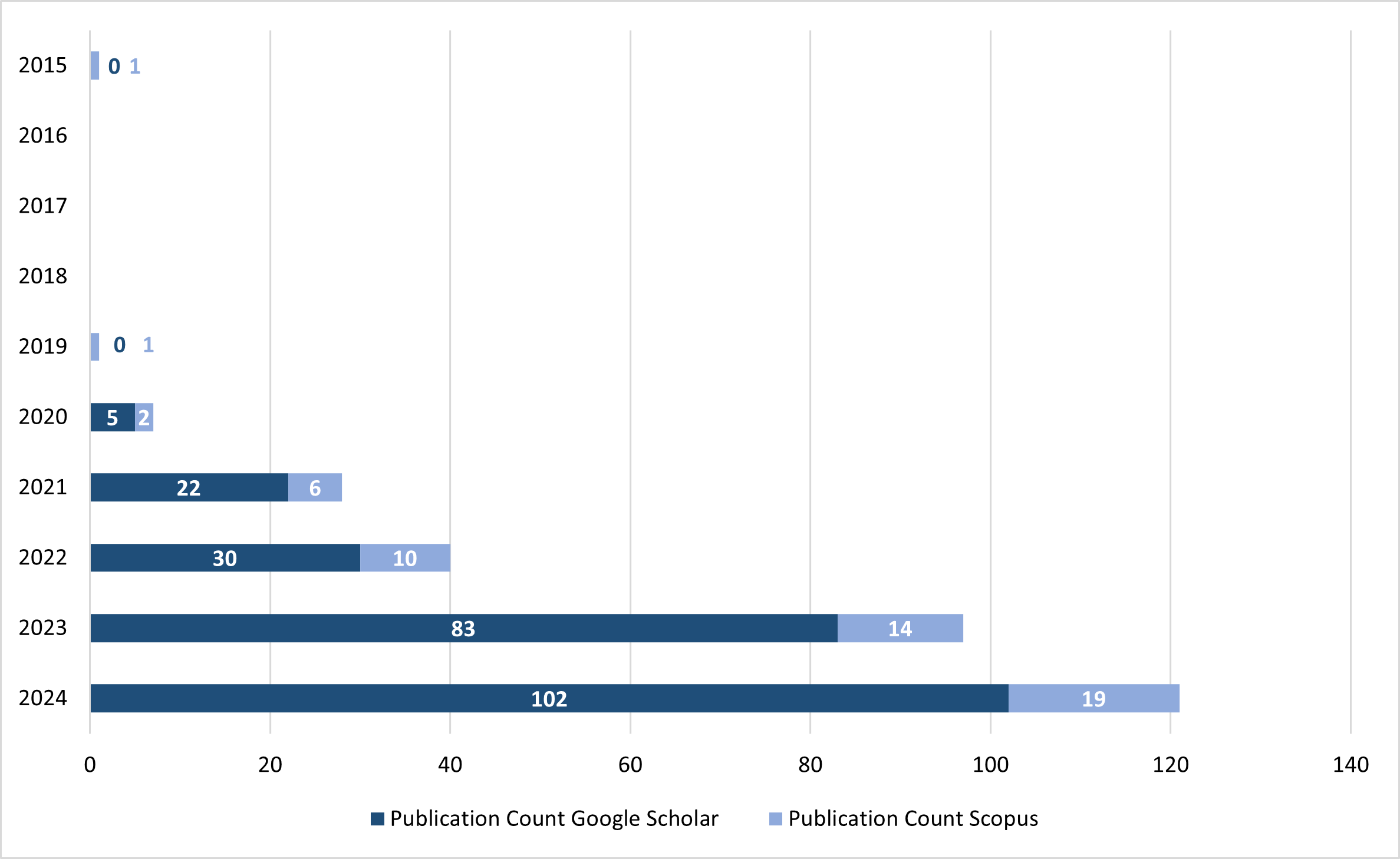}
    \label{fig:publications}
\end{figure}

This growing interest is shown in Fig. \ref{fig:publications}, which compiles the trend in QSOC publications over the years from two of the leading bibliographic databases, \textit{Scopus} and \textit{Google Scholar}. Specifically, the search was performed on publications that included the terms “\textit{Quantum Service-Oriented Computing}” and “\textit{Quantum Service}” and their possible variants in their content, and which fell within the subject area of computer science or engineering. In Fig. \ref{fig:publications}, the strong blue bars represent the publication counts from Google Scholar, while the light blue bars correspond to the Scopus search.

The trend in QSOC publications shows an initial phase of slow growth from 2015 to 2019, followed by a noticeable increase beginning in 2021 as the potential of quantum computing in service-oriented contexts gained traction. Starting in 2022, this growth became exponential, driven by key factors such as increased funding, industry adoption of hybrid quantum-classical systems, the inclusion of QSOC topics in major conferences, and the expanded availability of quantum hardware and platforms. And although 2015 is not yet over, the results obtained suggest that growth and interest in the subject continue.

From this analysis, we would like to highlight different works due to their importance for the scientific community in this field. One of the most cited papers in the field of quantum services is the 2024 paper by Nguyen et al. \cite{Nguyen2024}, entitled “\textit{QFaaS: A Serverless Function-as-a-Service Framework for Quantum Computing,}” published in Future Generation Computer Systems. The key contribution of this work is the introduction of QFaaS, a comprehensive, vendor-agnostic framework that leverages the serverless model to simplify and standardize the execution of quantum applications. The platform supports a wide range of quantum SDKs, simulators, and cloud providers, and incorporates an adaptive backend selection policy along with a caching mechanism to mitigate cold start delays, thus addressing practical issues that arise in dynamic quantum workloads.

Another notable contribution in the field is the 2020 paper titled "\textit{TOSCA4QC: Two Modeling Styles for TOSCA to Automate the Deployment and Orchestration of Quantum Applications}", presented at the 24th International Conference on Enterprise Distributed Object Computing (EDOC) \cite{Wild2020}. The primary contribution of this work is the proposal of two distinct modeling styles grounded in the TOSCA (Topology and Orchestration Specification for Cloud Applications) standard. These modeling approaches are specifically designed to automate the deployment and orchestration of quantum applications, enabling developers to define and manage quantum workloads in a standardized, declarative manner, paving the way for more scalable and manageable quantum software architectures.

On the other hand, one of the earliest efforts to explore the intersection between quantum computing and service-oriented architectures is the 2015 paper by Jatoth et al. \cite{Jatoth2015}, titled "\textit{QoS-Aware Web Service Composition Using Quantum-Inspired Particle Swarm Optimisation}", presented at the 7th KES International Conference on Intelligent Decision Technologies (KES-IDT 2015). In this work, the authors propose a novel approach to solving the QoS-aware web service composition problem using a quantum-inspired particle swarm optimization (QPSO) algorithm. This pioneering contribution laid the groundwork for future research by demonstrating how principles derived from quantum computing could be applied to enhance the performance of service composition mechanisms, particularly in multi-objective optimization contexts.

Finally, a more recent work, published in 2024, was entitled "\textit{A Reference Architecture for Quantum Computing as a Service}", published in the Journal of King Saud University – Computer and Information Sciences. In this work, Ahmad et al. \cite{Ahmad2024} propose a set of design guidelines and recommendations for building quantum-enabled microservices based on the quantum-classic split pattern. The paper’s key contributions include an empirically grounded reference architecture and a proof-of-concept implementation that demonstrates how quantum and classical components can be modularized into interoperable microservices. This work offers a practical blueprint for developing hybrid applications and contributes to the broader discourse on architecture patterns for Quantum Computing as a Service (QCaaS).
\section{Influence of the SE2030 Roadmap Workshop}
\label{sec:influence}


This paper is significantly influenced by the discussions and insights outlined during the "SE2030 Roadmap Workshop", a strategic event co-located with ACM SIGSOFT FSE on June 26-27, 2025, in Trondheim, Norway\footnote{2030 Software Engineering (2025). Retrieved from \url{https://conf.researchr.org/home/2030-se-2025}}. This gathered leading software engineering researchers to collectively define a vision for the discipline for the coming years. The workshop identified different foundational areas that will reshape software engineering: Agentic AI, AI for SE, SE for and by Humans, Sustainable SE, Quantum Software Engineering, among others. Each of these areas revealed tensions between established practices and the transformative impact of emerging technologies, including quantum computing.

The QSE discussions at the workshop recognized that quantum computing represents not just a technical evolution, but a change of paradigm that questions foundational assumptions of software design, architecture, and lifecycle management. One recurring theme was the necessity to rethink modularity and abstraction in light of quantum-specific constraints such as decoherence, entanglement, and probabilistic computation. While traditional software engineering relies on deterministic execution and well-defined state transitions, quantum software forces us to build systems under uncertainty, incomplete observability, and limited introspection capabilities. The SOQ model responds directly to this shift by proposing a service-oriented architecture that abstracts quantum resources as composable and discoverable services, shielding classical developers from the low-level mechanics of quantum logic gates and error models.

Participants recognized that while hybrid quantum-classical architectures are essential in the current NISQ era, the long-term vision for QSE should not be confined to hybrid systems. In contrast to models that treat quantum systems as auxiliary to classical workflows, SOQ aims to establish an architectural paradigm where quantum services can exist and operate independently, whether in quantum environments or in hybrid ecosystems. This independence is enabled by SOQ’s emphasis on loose coupling, platform-agnostic service definitions, and modular orchestration mechanisms. Several researchers at the workshop highlighted the need for robust intermediate layers, such as standardized quantum APIs, middleware, and orchestration frameworks, that not only bridge quantum and classical components when needed but also support fully quantum-native service ecosystems. These abstractions are central to SOQ’s identity, ensuring that quantum services remain portable, composable, and interoperable across both heterogeneous backends and diverse deployment contexts.

On the other hand, the discussions around AI for SE and Agentic AI revealed clear intersections with SOQ. For example, AI tools, such as Large Language Models (LLMs) and program synthesis engines, can be used to automate the generation of quantum circuits or hybrid service compositions. The opportunity to leverage Generative AI for designing SOQ service interfaces, configuring workflows, and generating test cases for probabilistic services was highlighted. At the same time, the validation of such systems presents novel challenges. As highlighted, the inherently non-deterministic behavior of quantum services requires new frameworks for correctness, trust, and confidence. The SOQ model can serve as a testing ground for advancing such verification models, especially in hybrid service contexts where observability and reproducibility are limited.

Additionally, the proposal for human-centric software engineering is particularly relevant to SOQ, which must navigate the tension between complex quantum principles and the goal of providing accessible, understandable interfaces to developers and domain experts. A topic of interest was the idea of ``quantum usability layers'' that translate quantum operations into domain-level constructs, such as finance, logistics, or chemistry, facilitating broader adoption. SOQ can offer a pathway for implementing these usability layers as modular services that encapsulate domain-specific quantum capabilities (e.g., optimization, simulation) without requiring users to understand the underlying hardware or algorithms.

On the sustainability front, the energy-intensive nature of quantum computing was flagged as a key concern. While quantum algorithms may reduce the time complexity of certain tasks, the overhead of cooling quantum processors, running redundant shots, and maintaining coherence across qubits can offset these benefits. In this regard, SOQ can act as a governance and optimization layer, managing resource allocation and execution policies based on sustainability metrics (e.g., energy per useful operation, carbon footprint per service call, etc.). This could enable future research on green quantum computing, where orchestration policies account for both performance and ecological impact.

Ultimately, the SE2030 workshop underscored the need for shared frameworks, standards, and open toolchains that allow the SE community to engage meaningfully with quantum computing. The SOQ vision is a tangible response to this demand, it encapsulates quantum capabilities within a familiar and extensible architectural model, enables hybridization through well-understood service composition patterns, and promotes research into new models of validation, security, and sustainability. 

In short, all these observations and discussions that emerged during the SE2030 workshop strongly resonated with the authors' current work on this paper, motivating us to explicitly position SOQ as a first-class service-oriented paradigm for quantum computing, going beyond the previous concept of QSOC. This paper incorporates many of the challenges articulated in the workshop, which are detailed and expanded upon in the following sections of this manuscript.

\section{Challenges in Service-Oriented Quantum}
\label{sec:challenges}
The SOQ paradigm leverages the principles of SOC to the quantum domain, ensuring modularity, scalability, and interoperability in hybrid quantum-classical environments. Unlike previous approaches that treated quantum services as adaptations within SOC, SOQ positions quantum computing as a first-class participant in service-oriented ecosystems. This shift enables quantum functionalities to be exposed as independent, loosely coupled services that can be composed, orchestrated, and integrated seamlessly with classical IT infrastructures.

However, several challenges must be addressed to fully realize the benefits of SOQ. These include the need for standardized quantum service interfaces, efficient quantum-classical communication protocols, and adaptive orchestration mechanisms that can handle the dynamic nature of quantum computations. Furthermore, given the limitations of NISQ devices, SOQ must accommodate error-prone quantum services while ensuring fault tolerance and reliability in hybrid workflows. Addressing these challenges will be essential to developing a robust and efficient Quantum-as-a-Service (QaaS) ecosystem, ultimately enabling broader adoption of quantum technologies across various industries.

\subsection{Relevant works that address the challenges}
\label{subsec:relevant_works_challenges}

\begin{figure}[!ht]
\centering
   \caption{Challenges, possible use cases, and relevant works addressing the challenges.}
    \includegraphics[width=1\textwidth]{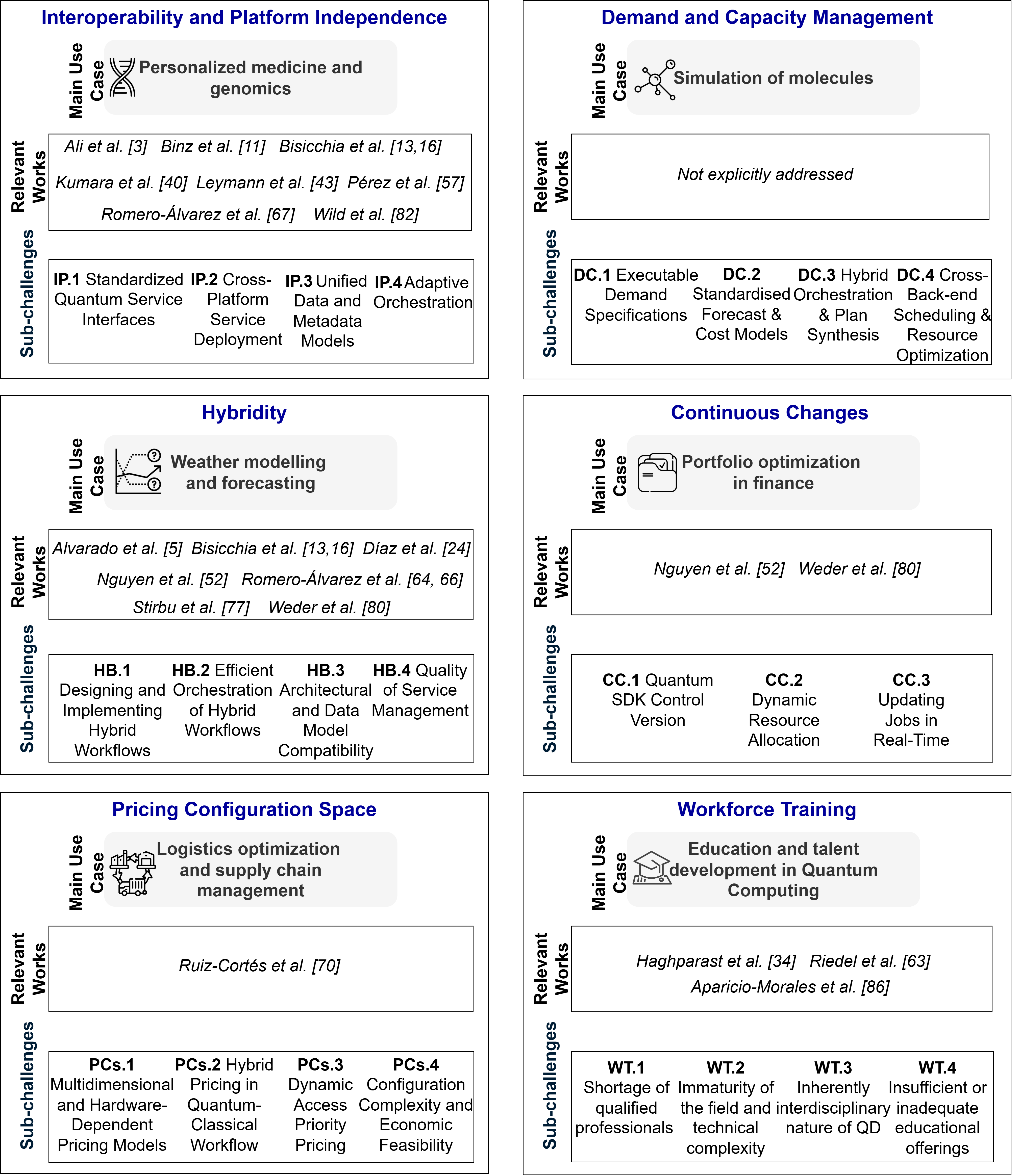}
    \label{fig:qsoc_challenges}
\end{figure}

The diverse research efforts addressing different challenges and aspects, from architecture design to practical implementations, reflected the growing interest in this area, as seen before. This section reviews some of the most relevant contributions in this field that seek to solve the problems and challenges detected. Fig.  \ref{fig:qsoc_challenges} shows a correlation between SOQ challenges, sub-challenges, the use cases raised, and the most relevant research works.  As far as we know, there are still challenges that have not yet been addressed, which reaffirms the crucial need for future work and studies in this area.

Some of the earliest works exploring the intersection of SOC and quantum computing are by Karoline et al. \cite{Wild2020}, where the authors propose several modelling styles based on the TOSCA standard. They discuss the challenges of orchestrating and deploying quantum and classical services, emphasizing the need for interoperability mechanisms. Moreover, in Kumara et al. \cite{Kumara2021}, the authors present a model-driven methodology based on QSOC that allows developers to build hybrid enterprise applications collaboratively. 

In studies focused on the coexistence of classical and quantum services, the work of Ali and Yue \cite{Ali2023} stands out, where they emphasize the need for a quantum-oriented paradigm to address challenges and highlight key issues such as interoperability, abstraction mechanisms, and platform independence. Moreover, in Romero et al. \cite{Romero2024c}, a scheduler is proposed to execute quantum circuits on cloud service providers. Similarly, Alvarado et al. \cite{Alvarado2022,alvarado2023quantum} present a guide to converting quantum circuits into web services. Furthermore, the study by Nguyen et al. \cite{Nguyen2024} introduces the concept of QFaaS, leveraging the serverless model for the execution of quantum circuits.

Orchestration is a key challenge in SOQ, as quantum workflows require sophisticated middleware solutions to manage heterogeneous quantum back-ends. Some works address this issue by proposing orchestration frameworks for quantum applications, which enable workflow automation and efficient resource allocation \cite{Leymann2019,Weder2020}. 

An alternative approach to managing hybrid quantum-classical workflows, while also improving interoperability, focuses on distributing the quantum workload on a \textit{shot-by-shot} basis, rather than sending an entire quantum job to a single QPU. This idea is explored in the work by Bisicchia \textit{et al.} \cite{Bisicchia2024c,Bisicchia2023} in which the authors introduce a \textit{Quantum Broker} that automatically splits the requested number of shots among multiple QPUs according to user-customizable policies.  Allocating subsets of the total shots to different QPUs improves both flexibility and resilience. If one QPU experiences downtime or failure, its share of shots can be reallocated to functioning machines, and only a fraction of the overall results are affected. 

Similarly, DevOps methodologies have been adapted to quantum computing, incorporating automated deployment \cite{Binz2013}, containerization \cite{Stirbu2024}, and delivery practices to ensure software reliability in hybrid quantum-classical environments \cite{Romero2024,alvarado2023devops}. These studies emphasize the need for automated testing and validation of quantum circuits to mitigate the instability of quantum hardware.

Regarding quantum service deployment, recent research has also explored the modularization of hybrid information systems to ensure platform compatibility \cite{Perez2022}. Other studies have investigated testing implications \cite{Muqeet2024} and the quality of deploying quantum services \cite{Diaz2025} in cloud environments, proposing best practices to test quantum circuits.
 
Collectively, these works establish the interest, the need to integrate quantum and classical systems, and the technical feasibility, but above all, more efforts are needed in this area. Therefore, based on the current work of the community and the authors' experience in this field over recent years, several challenges have been identified. The main ones are as follows.

\subsection{Interoperability and Platform Independence}
\label{subsec:interoperability_and_platform_independence}

Interoperability and platform independence stand as foundational pillars for realizing the Service-Oriented Quantum paradigm. Without robust mechanisms to integrate heterogeneous quantum and classical resources, the promise of scalable, reusable, and maintainable quantum-enabled services remains unattainable. Yet the present quantum computing landscape is characterized by severe fragmentation \cite{Zhao2020}, the risk of vendor lock-in, and rapidly evolving software stacks, factors that continue to impede the emergence of open, flexible, and accessible quantum infrastructures.

The ecosystem is fragmented mainly due to a tendency by quantum providers to develop their own QPUs, software stacks, SDKs, and languages \cite{Serrano2022}. This diversity has generated a proliferation of proprietary interfaces and non-standardized circuit representations, resulting in programming models and toolchains that are usually challenging to integrate across platforms. The absence of standardized abstractions manifests in several problematic ways \cite{Bisicchia2024}. Algorithms, workflows, or services created for one provider are rarely portable to others without substantial manual re-engineering, which undermines the vision of reusable, cross-platform quantum software. Users are often confined within a single vendor’s ecosystem, restricting their capacity to select the most suitable hardware for a given problem and, in turn, stifling both competition and innovation. Moreover, the research community suffers from duplicated efforts and missed opportunities for synergy, as solutions developed in one stack are typically not composable with those from another, fragmenting progress and limiting large-scale adoption.

Addressing these challenges, the community has started to pursue standard intermediate representations and interoperability layers. Solutions such as OpenQASM \cite{Cross2022}, the Quantum Intermediate Representation (QIR) \cite{Luo2025}, and toolkits like \textit{(t|ket⟩)} \cite{Sivarajah2020} and XACC \cite{Mccaskey2020} represent promising steps toward enabling quantum circuit portability across back-ends. Provider-neutral SDKs, such as qBraid\footnote{\url{https://github.com/qBraid}} and QuantumExecutor \cite{Bisicchia2025}, attempt to abstract away some of the provider-specific complexities. Nevertheless, progress is uneven. Even when nominal compatibility exists, differences in supported gate sets, qubit connectivity, and error models may introduce semantic mismatches, which can degrade performance or even yield incorrect results. Furthermore, because the execution of quantum services remains closely tied to the physical characteristics of specific QPUs (such as their topology, coherence times, and susceptibility to noise) true platform independence remains an elusive goal.

A promising strategy to address both interoperability and platform independence is the introduction of meta-platforms or “Virtual Quantum Providers” (VQPs) \cite{Bisicchia2024}. These act as intermediaries that abstract over the diversity of providers, exposing unified interfaces and automating provider selection, resource management, and job routing. Pioneering examples include qBraid, Classiq\footnote{\url{https://www.classiq.io}}, PlanQK \cite{Falkenthal2024}, and orchestration frameworks that support multi-stack software pipelines.

These VQPs provide several key advantages:
\begin{itemize}
    \item Unified API: Developers can target a single programming interface, while the VQP handles translation, compilation, and optimization for multiple hardware targets.
    
    \item Automated provider selection: Algorithms for cost, performance, or availability can dynamically select the best QPU for each task, maximizing resource utilization and minimizing queue times \cite{Bisicchia2023,Bisicchia2023b}.
    
    \item Reduced vendor lock-in: Decoupling application logic from specific providers reduces migration costs and future-proofs software investments.
    
    \item Composable pipelines: Multi-stack workflows become feasible, allowing compilation, simulation, and execution stages to leverage the unique strengths of various SDKs and providers in a seamless fashion \cite{Bisicchia2024c}.
\end{itemize}

Yet, challenges remain. VQPs must cope with evolving provider APIs, changing hardware capabilities, and the ongoing lack of robust, community-wide standards. The abstraction gap can mask, but not fully eliminate, semantic mismatches and performance inconsistencies rooted in the physical layer.

Building on these insights, we identify several concrete research challenges to advancing interoperability and platform independence within SOQ architectures (Table \ref{tab:interop}).

\begin{table}[!ht]
\caption{Interoperability and Platform Independence Challenges for SOQ.}
\centering
\begin{tikzpicture}
\node (table) [inner sep=0pt] {
\begin{tabular}{c|p{4.3cm}|p{8cm}}
    \rowcolor{blue!10}
    \textbf{\#} & \textbf{Challenge} & \textbf{Description} \\
    \hline
    \textbf{IP.1} & Standardized Quantum Service Interfaces & Define and formalize universal, provider-agnostic interfaces for quantum services, including inputs/outputs, resource requirements, and quality attributes. \\
    \hline
    \textbf{IP.2} & Cross-Platform Service Deployment & Develop abstraction and deployment mechanisms enabling seamless execution of quantum services across diverse providers. \\
    \hline
    \textbf{IP.3} & Unified Data and Metadata Models & Create interoperable data formats for quantum-classical exchange, along with standardized metadata to describe job context, fidelity, and provenance. \\
    \hline
    \textbf{IP.4} & Adaptive Orchestration & Implement orchestration engines capable of real-time provider selection, resource negotiation, and workflow optimization based on availability and performance demands. \\
\end{tabular}

};
\draw [rounded corners=.5em] (table.north west) rectangle (table.south east);
\end{tikzpicture}
\label{tab:interop}
\end{table}

\begin{description}
\item[IP.1 Standardized Quantum Service Interfaces]
\mbox{ }
\\
A fundamental step toward interoperability is the establishment of standardized, universal service definitions-analogous to GraphQL \cite{Quinamera2023}, REST or OAS in classical SOC. This involves specifying canonical types for quantum inputs (e.g., circuits, oracles, parameterized gates), outputs (bitstrings, state vectors, error estimates), and service metadata (back-end topology, calibration data). Such standardization would allow developers to compose and reuse services independent of the underlying hardware or SDK, while supporting the automatic negotiation of resource needs and QoS expectations \cite{Bisicchia2023b}.
\\
\item[IP.2 Cross-Platform Service Deployment]
\mbox{ }
\\
Realizing platform independence requires not only standard interfaces, but also deployment mechanisms that enable quantum services to run “anywhere”. Techniques such as intermediate circuit representation, automatic transpilation, and runtime adaptation to target-specific constraints (e.g., gate sets, qubit topology) are needed. Existing compilers like \textit{(t|ket⟩)} and frameworks like ProjectQ provide some of this flexibility, but further work is required to make these solutions transparent, robust, and integrated within the SOQ lifecycle.
\\
\item[IP.3 Unified Data and Metadata Models]
\mbox{ }
\\
A neglected but critical aspect is the standardization of data models for quantum-classical exchange. Measurement results, error rates, and calibration data must be represented in forms that are compatible and easily consumable by classical services, with rich metadata to support traceability, reproducibility, and statistical analysis. Without such standards, hybrid workflows will remain brittle and error-prone.
\\
\item[IP.4 Adaptive Orchestration]
\mbox{ }
\\
Future SOQ platforms may include orchestration engines that monitor hardware status, predict queue times and error rates, and dynamically re-route jobs to the most appropriate provider \cite{Bisicchia2023}. This requires integration of hardware monitoring APIs, predictive analytics, and policy-driven resource allocation \cite{Bisicchia2023b}. The end goal is “liquid” quantum services, adaptively migrating, scaling, and optimizing execution based on a continuously changing environment.

\end{description}

\noindent
While the push for interoperability and platform independence is widely supported, a well-informed skeptic might argue that too much abstraction risks diluting performance, obfuscating critical hardware-specific optimizations, and creating a “lowest common denominator” effect that undermines practical utility. Furthermore, as the field is rapidly evolving, premature standardization could lock the ecosystem into suboptimal patterns. It may be argued that targeted, use-case-specific pipelines, rather than universal abstractions, will remain necessary for some time, especially in high-stakes applications where performance or reliability is paramount.

Nonetheless, history in classical computing suggests that the long-term benefits of interoperability (such as reduced duplication, improved reproducibility, faster innovation cycles) outweigh the temporary losses from abstraction. The challenge for SOQ is to balance these trade-offs: enabling portability and composability without sacrificing the ability to leverage unique quantum hardware capabilities.

Interoperability and platform independence are indispensable for the maturation of quantum software engineering and the realization of SOQ. Overcoming the current landscape of fragmentation, vendor lock-in, and brittle software stacks will require sustained efforts in standardization, orchestration, and community collaboration. Virtual Quantum Providers, unified service interfaces, and adaptive orchestration engines represent promising paths forward, but require careful design to avoid masking fundamental performance or reliability gaps. Ultimately, addressing these challenges is not merely a technical necessity, it is an enabler for the practical impact of quantum computing across scientific, industrial, and societal domains. A unified, interoperable SOQ layer would allow seamless migration, composition, and scaling, accelerating discovery and democratizing access to cutting-edge computational resources.
\subsection{Demand and Capacity Management}
\label{subsec:demand_capacity_management}

Typically, quantum computing relies on best-effort runs on a single back-end: circuits are executed multiple times (called \textit{shots}), queued by providers, and in hybrid workflows interleaved with classical computation \cite{Nielsen2010}. Service-Oriented Quantum aims to make resource management and service time explicit, to guide the planning and execution of quantum workloads across heterogeneous providers.

Quantum computing resources remain scarce and expensive. Unlike classical cloud computing, where resource allocation and autoscaling are well understood, quantum capacity face hardware-specific constraints and rapidly changing operational conditions. In particular, \emph{service times}, \emph{gate fidelities}, and \emph{qubit availability} jointly determine the \emph{effective} throughput of a quantum service \cite{Zhao2020}. Queue-based execution and heterogeneous pricing introduce cost-latency trade-offs that demand explicit management, while hybrid jobs add classical billing to the mix. Adapting demand-management patterns from microservice architectures to this context calls for new abstractions for quality-aware scheduling, budget control, and quantum-classical co-scheduling across \emph{heterogeneous} devices and providers.

SOQ aims at making these levers explicit and actionable. End-to-end performance is governed by the \emph{service time} budget
\[
T_{\mathrm{svc}} \;=\; T_{\mathrm{queue}} \;+\; T_{\mathrm{quantum}} \;+\; T_{\mathrm{classical}}
\]
where $T_{\mathrm{queue}}$ reflects provider scheduling and concurrency caps, $T_{\mathrm{quantum}}$ is the scheduled duration across all shots (including measurement and idle waits due to routing or synchronization), while $T_{\mathrm{classical}}$ covers optimization, data movement, as well as pre- and post-processing in hybrid loops. At the same time, current NISQ devices are affected by decoherence, gate errors, and readout inaccuracies that distort ideal evolutions and output distributions, lowering computational fidelity and complicating the separation of correct from spurious results. Effective demand and capacity management in SOQ is thus essential for optimizing resource allocation, ensuring efficient quantum-classical task scheduling, and mitigating the high costs of quantum processing.

Effective planning, therefore, must treat fidelity, service time, and qubit availability as first-class controls: schedule submissions when calibration is favorable, route circuits to qubit pairs with higher two-qubit fidelities that are currently enabled, and, when feasible, distribute shots across eligible back-ends to reduce wall-clock time and hedge against transient noise. Pricing choices are inseparable from these decisions: large, precision-driven shot portfolios with tight SLAs may justify reservations, and exploratory runs often benefit from on-demand and dynamically priced options, paying only for the shots and tasks actually executed. 
To make such decisions reproducible and auditable, SOQ services should expose \emph{capability contracts} that an orchestrator enforces at admission and uses to steer compilation, routing, scheduling, and measurement planning. This \emph{capability-based programming} style allows requesters to express priorities over \emph{heterogeneous} resources (e.g., preferring calibrated qubit pairs or bounding queue delays) and to align execution with domain goals.

The current research on capacity management focuses predominantly on QPU selection by optimizing metrics over devices, circuits, and environment (e.g., queue length, cost, availability). Garcia-Alonso et al. \cite{Garcia2021} select QPUs given architecture and circuit width while letting users prioritize speed or cost. Marie et al. \cite{Salm2020,Salm2022} go further by selecting both a circuit implementation and a QPU based on width, depth, SDK compatibility, and input-dependent rules, with extensions for compiler comparison and ML-based pruning/optimization. Other proposals jointly consider estimated output fidelity and queue waiting time in adaptive schedulers \cite{Ravi2021}, integrate quantum within enterprise cloud selection using static attributes such as qubit count and queue length \cite{Grossi2021}, or predict the best combination of QPUs, compilers, and compilation parameters to maximize execution fidelity \cite{Quetschlich2023}. Despite these advances, most approaches still choose a \emph{single} best QPU (or QPU--circuit pair) and execute all measurement shots there. By contrast, \emph{shot-wise distribution} \cite{Bisicchia2024b,Bisicchia2023} exploits the statistical independence of shots to distribute them across multiple \emph{heterogeneous} QPUs, enabling parallelism, resilience to device drift, and adaptive load balancing, an approach that aligns naturally with SOQ’s governance of service time, fidelity, and price.

Building on the above discussion, we identify four concrete research challenges that advance demand and capacity management within SOQ (Table \ref{tab:dc-soq}). These challenges translate the execution model (shots, queues, hybrid loops), service-time budgeting, hardware quality (fidelity and qubit availability), and pricing choices into actionable design problems for governed, repeatable operation across \emph{heterogeneous} devices and providers.

\begin{table}[!ht]
\caption{Demand and Capacity Management Challenges for SOQ.}
\centering
\begin{tikzpicture}
\node (table) [inner sep=0pt] {
\begin{tabular}{c|p{4.3cm}|p{8cm}}
    \rowcolor{blue!10}
    \textbf{\#} & \textbf{Challenge} & \textbf{Description} \\
    \hline
    \textbf{DC.1} & Executable Demand Specifications & User-facing, provider-agnostic ways to express execution priorities and constraints (e.g., latency targets, spend ceilings, fidelity targets, eligible devices, maximum shots, reservation requirements) that an orchestrator can validate, audit, and enforce. \\
    \hline
    \textbf{DC.2} & Standardized Forecast \& Cost Models & Cross-provider interfaces exposing forecastable service attributes, queue delay, run time, price, and expected fidelity, with uncertainty bands and what-if queries, enabling apples-to-apples planning and pre-execution quotes. \\
    \hline
    \textbf{DC.3} & Hybrid Orchestration \& Plan Synthesis & Orchestrators that derive end-to-end execution plans consistent with user specifications: selecting devices and pricing modes, sizing and ordering shot batches, aligning classical steps, and preparing distribution strategies across \emph{heterogeneous} back-ends. \\
    \hline
    \textbf{DC.4} & Cross-Backend Scheduling \& Resource Optimization & Scheduling policies and algorithms that allocate and rebalance workload (including shot-wise distribution) to optimize cost/latency under constraints, while coping with concurrency caps, calibration changes, and multi-tenant fairness. \\
\end{tabular}
};
\draw [rounded corners=.5em] (table.north west) rectangle (table.south east);
\end{tikzpicture}
\label{tab:dc-soq}
\end{table}

\begin{description}
\item[DC.1 Executable Demand Specifications]
\mbox{ }\\
Provide a capability-based, machine-checkable specification through which users declare what the service must achieve and under which bounds. Typical fields include service time, SLA targets, price ceilings, maximum shots, preferred pricing mode (on-demand or reserved), fidelity targets, and eligible devices/providers. Specifications should support composition (pipelines of circuits and hybrid loops), be statically validated before admission, and remain auditable during execution so that decisions (e.g., rerouting or early stopping) can be traced back to declared constraints.
\\
\item[DC.2 Standardised Forecast \& Cost Models]
\mbox{ }\\
Define provider interfaces that expose pre-execution forecasts for queue time, quantum run time (per circuit and per shot), classical overhead for hybrid steps, expected monetary cost (per-shot/per-task and reservation rates), and expected computation fidelity. These interfaces should return point estimates with confidence intervals, specify the forecasting horizon and assumptions (e.g., current queue state), and support what-if queries (e.g., “50k shots across two back-ends”). Standardization enables planners to compare options across providers and to produce consistent, auditable quotes prior to submission.
\\
\item[DC.3 Hybrid Orchestration \& Plan Synthesis]
\mbox{ }\\
Develop orchestration services that translate user specifications and provider forecasts into executable plans spanning quantum and classical resources. Plans should decide when to submit (or reserve), which back-ends to target, how to partition and order shot batches, how to align classical optimization steps with expected device availability, and whether to employ heterogeneous distribution. They must include rollback and adaptation rules (e.g., switch device if availability changes) and produce artifacts for governance: a rationale, predicted $T_{\mathrm{svc}}$ and spend, and checkpoints for mid-course corrections.
\\
\item[DC.4 Cross-Back-end Scheduling \& Resource Optimization]
\mbox{ }\\
Design schedulers that implement the plan under real-time conditions, allocating jobs and shots across multiple eligible back-ends to minimize cost and latency while respecting constraints from DC.1. Key capabilities include queue-aware admission control, fidelity-aware routing and compilation choices, adaptive shot allocation with early stopping, and on-the-fly rebalancing when queue states, calibration outcomes, or availability change. The scheduler should support multi-tenant fairness and provide utilization metrics that feed back into forecasting and future plans.
\end{description}

Taken together, these elements point to SOQ as a practical pathway from best-effort runs to predictable, cost-aware quantum services. It achieves this by making critical metrics, such as service time, gate fidelity, and qubit availability, explicit and by capturing user intent through executable demand specifications. In addition, SOQ exposes provider forecasts and cost models, integrates hybrid orchestration, and enables cross-back-end scheduling, including shot-wise distribution. Through this combination, SOQ aligns pricing and latency with accuracy objectives across heterogeneous providers. The outcome is a repeatable and auditable execution process that mitigates scarcity and noise, improves resource utilization, and allows domain teams to meet service-level agreements without compromising scientific or industrial objectives.
\subsection{Hybridity}
\label{subsec:hybridity}

Hybrid quantum-classical workflows require efficient coordination between systems. Many quantum algorithms rely on classical pre-processing and post-processing steps, creating potential bottlenecks in service orchestration.

Recent research efforts have started to address some of these challenges by proposing early architectures and runtime environments for hybrid execution. Examples include AWS Hybrid Jobs and IBM Quantum Runtime, which provide frameworks for combining quantum and classical computations in a coordinated way, and the Qubernetes platform \cite{Stirbu2024}, which explores cloud-native execution for hybrid quantum-classical workflows. The natural emergence of these hybrid workflows implies a design in which it is necessary to identify relevant aspects of Quality of Service (QoS) by achieving a good balance among the non-functional properties of preexisting components or services, such as reliability, execution time, and cost. Consequently, it is necessary to analyze the trade-offs within the design space. 

In this context, it is essential to consider the constraints and uncertainties of classical service components, as well as the additional constraints that naturally arise from integrating quantum-classical service-based systems. For instance, quantum services can only be deployed on specific hardware, come with different properties, and follow different pricing models. 
Moreover, these services introduce additional sources of uncertainty, such as the non-deterministic output of quantum circuits. Therefore, it is necessary to develop new tools, methodologies, or frameworks that facilitate the correct distribution and coordination of services across the IT infrastructure to ensure the desired quality of service, which can be very critical, as in the case of software systems.

Recent scientific work has begun to address these challenges. Cranganore et al. \cite{Cranganore2024} propose hybrid cloud architectures for scientific workflows, discussing scheduling and resource allocation trade-offs; O’Riordan et al. \cite{ORiordan2020} explore hybrid workflows for natural language processing; and Sivarajah et al. \cite{Sivarajah2020} develop a retargetable compiler \textit{(t|ket⟩)} that facilitates hybrid execution by bridging quantum programs with heterogeneous back-ends. These contributions mark an important step forward, yet they also underscore how immature the current methodologies are and how much remains to be done to generalize these approaches beyond isolated use cases.

Building on these insights, in Table \ref{tab:hibridity} we identify several forward-looking research challenges that will become increasingly relevant in the coming years, as hybrid quantum-classical systems move from experimental setups to production-level environments.

\begin{table}[!ht]
\caption{Hybrid Quantum-Classical Challenges for SOQ.}
\centering
\begin{tikzpicture}
\node (table) [inner sep=0pt] {
\begin{tabular}{c|p{4.3cm}|p{8cm}}
    \rowcolor{blue!10}
    \textbf{\#} & \textbf{Challenge} & \textbf{Description} \\
    \hline
    \textbf{HB.1} & Designing and Implementing Hybrid Workflows. & Methods and tools to model, implement, and validate hybrid quantum-classical workflows effectively. \\
    \hline
    \textbf{HB.2} & Efficient Orchestration of Hybrid Workflows. & Dynamic and fault-tolerant orchestration mechanisms for hybrid computations. \\ 
    \hline
    \textbf{HB.3} & Architectural and Data Model Compatibility. & Ensuring consistent and interoperable architectures and data formats across quantum and classical components. \\ 
    \hline
    \textbf{HB.4} & Quality of Service Management in Hybrid Environments. & Defining and enforcing QoS metrics and SLAs specific to hybrid quantum-classical workflows.\\ 
\end{tabular}
};
\draw [rounded corners=.5em] (table.north west) rectangle (table.south east);
\end{tikzpicture}
\label{tab:hibridity}
\end{table}


\begin{description}
\item[HB.1 Designing and Implementing Hybrid Workflows]
\mbox{ }
\\
Designing and implementing hybrid workflows is a foundational challenge for hybrid quantum-classical computing. Current implementations are often handcrafted, domain-specific, and lack formal design patterns or reusable templates. This hinders scalability and makes it difficult to adapt workflows to different contexts or hardware. To overcome this, the community needs to develop systematic methodologies and domain-specific languages that allow engineers to express hybrid workflows at a high level of abstraction, along with hybrid-aware validation and testing strategies. 

Steps forward include defining formal workflow models, creating repositories of reusable hybrid templates, and integrating hybrid workflows into continuous integration and deployment pipelines to facilitate iterative development and testing.

One notable example is in climate modeling and weather forecasting, where hybrid workflows are already proving valuable. Since these simulations require vast computational power and information obtained from classical data sources, managing when and how computational infrastructure is utilized is crucial to avoid bottlenecks and ensure critical computations are completed within tight time constraints. Google’s Analog-Digital Hybrid Quantum Simulation\footnote{IEEE Spectrum (2023). Google’s Analog-Digital Hybrid Quantum Simulation. Retrieved from \url{https://spectrum.ieee.org/quantum-simulation}} has demonstrated how quantum subroutines can enhance long-term climate predictions when orchestrated within large classical simulation pipelines. In such contexts, designing reusable and verifiable hybrid workflows is critical to ensuring reliable predictions that can inform disaster response planning.
\\
\item[HB.2 Efficient Orchestration of Hybrid Workflows]
\mbox{ }
\\
Efficient orchestration is critical for dynamic allocation of tasks between quantum and classical resources in a way that minimizes latency, maximizes throughput, and maintains fault tolerance. Future research should focus on designing hybrid orchestration policies that explicitly consider performance uncertainty and resource heterogeneity. 

Promising steps include using AI-based schedulers to predict quantum back-end availability, adding mechanisms to orchestrate classical and quantum parts, and developing hybrid workflow engines that optimize task allocation in real time.

BMW\footnote{BMW Group (2025). Quantum Computing News. Retrieved from \url{https://www.bmwgroup.com/en/news/general/2025/quantum-computing.html}} and Airbus\footnote{Airbus (2024). Airbus Quantum Computing Challenge. Retrieved from \url{https://www.airbus.com/en/innovation/digital-transformation/quantum-technologies/airbus-and-bmw-quantum-computing-challenge}} provides an illustrative case, that are experimenting with hybrid workflows for aerodynamics optimization and supply chain logistics. In these contexts, orchestrators must determine how and when to offload computationally expensive subproblems to quantum processors while ensuring that the classic workflow structure remains efficient and synchronized.
\\
\item[HB.3 Architectural and Data Model Compatibility]
\mbox{ }
\\
The incompatibility of the architectural and data model between quantum and classical components remains a significant barrier to seamless integration. Classical systems use deterministic floating-point representations, while quantum results are inherently probabilistic and often represented as amplitudes or density matrices. Future work should prioritize the development of standard intermediate representations and hybrid architecture reference models, along with automated translators that convert quantum outputs into formats that classical components can process efficiently.

An illustrative example is in hybrid molecular dynamics simulations \cite{Irie2021}, where quantum electronic structure calculations produce data that must be integrated into classical force-field models for biomolecular interactions. Bridging this gap effectively could accelerate drug discovery pipelines by making hybrid simulations more robust and easier to implement across platforms. Moreover, IoT edge devices collect data continuously, and quantum solvers in the cloud can optimize routes and inventories \cite{Peelam2024}. Determining the optimal balance between local pre-processing at the edge and remote quantum computation is critical to ensuring timely, reliable decisions.
\\
\item[HB.4 Quality of Service Management in Hybrid Environments]
\mbox{ }
\\
Finally, managing QoS in hybrid workflows is a critical but complex challenge, arising from the fundamentally different performance, reliability, and cost profiles of quantum and classical services. Classical components are generally predictable, with mature monitoring and SLA mechanisms, whereas quantum components exhibit stochastic behavior, variable execution times, and hardware-specific error rates that can fluctuate depending on calibration and environmental conditions. This asymmetry creates difficulties in defining, negotiating, and enforcing meaningful QoS guarantees for hybrid applications.

Addressing this challenge requires several coordinated efforts. First, hybrid-specific QoS metrics must be defined to accurately capture the interplay between classical throughput and quantum probabilistic fidelity. Such metrics should reflect not only standard properties like response time and availability, but also quantum-specific aspects such as success probability, decoherence impact, and number of repetitions (shots) required to achieve statistical significance. Second, SLA negotiation protocols need to evolve to account for hybrid dependencies, allowing contracts to specify acceptable ranges of performance that consider quantum uncertainty. Finally, real-time hybrid monitoring infrastructures must be developed to continuously track and analyze QoS indicators across the heterogeneous environment, detecting anomalies and providing actionable feedback to orchestrators and end-users.

The impact of this challenge is evident in domains where regulatory or mission-critical constraints demand consistent performance. Financial risk analysis systems, for example, depend on predictable response times and accuracy for compliance reporting. Hybrid Monte Carlo simulations\footnote{Google Research (2022). Hybrid Quantum Algorithms for Quantum Monte Carlo. Retrieved from \url{https://research.google/blog/hybrid-quantum-algorithms-for-quantum-monte-carlo}} that combine classical and quantum resources must meet strict SLAs while balancing cost and resource utilization, requiring hybrid-aware SLA management to avoid regulatory breaches or costly re-runs. 

\end{description}

\subsection{Continuous changes}
\label{subsec:continuous_changes}

Quantum services are undergoing continuous evolution, characterized by frequent modifications in pricing models, availability, and feature sets. This dynamic nature poses significant challenges for the long-term planning, contracting, and orchestration of quantum services. Cloud providers frequently introduce new QPUs, necessitating continuous updates in service compositions to ensure compatibility and optimal performance. A notable example is Amazon Braket, which grants developers access to different quantum computing technologies from multiple hardware providers, such as IonQ, D-Wave, or Rigetti. However, due to the constant evolution of the quantum ecosystem, certain machines become unavailable over time. In addition, some services are no longer available directly from quantum providers, but are instead offered through alternative services, as seen in the case of D-Wave\footnote{AWS Quantum Technologies Blog (2022). Using D-Wave Leap from the AWS Marketplace. Retrieved from \href{https://aws.amazon.com/es/blogs/quantum-computing/using-d-wave -leap-from-the-aws-marketplace-with-amazon-braket-notebooks-and-braket-sdk/} {https://aws.amazon.com/es/qc-dwave-marketplace}}, where customers can no longer access the D-Wave 2000Q and Advantage systems through Amazon Braket.

Similarly, IBM Quantum has introduced a series of advancements in its quantum computing services\footnote{IBM Quantum Roadmap. Retrieved from \url{https://www.ibm.com/roadmaps/quantum}}, including the deployment of new quantum machines and frequent updates to its quantum programming language. These continuous modifications add further complexity to the development and maintenance of quantum service-oriented architectures, requiring adaptive solutions for interoperability and orchestration. In this context, techniques for monitoring service quality and ensuring seamless transitions between quantum hardware generations are crucial to maintaining service reliability and usability. Therefore, without standardized quantum-classical communication protocols and the continuous changes, the integration of quantum solvers into existing financial platforms would require recurrent customizations. Table \ref{tab:continuouschanges} summarizes some of these challenges that have been identified.

\begin{table}[!ht]
\caption{Continuous Changes Challenges for SOQ.}
\centering
\begin{tikzpicture}
\node (table) [inner sep=0pt] {
\begin{tabular}{c|p{4.3cm}|p{8cm}}
    \rowcolor{blue!10}
    \textbf{\#} & \textbf{Challenge} & \textbf{Description} \\
    \hline
    \textbf{CC.1} & Quantum SDK Control Version. & Tool to facilitate the migration of different providers' SDKs. \\
    \hline
    \textbf{CC.2} & Dynamic resource allocation & Quantum computing efficiency relies on ``liquid'' applications that adapt across the quantum-classical continuum, optimizing resources and accelerating practical solutions.\\ 
    \hline
    \textbf{CC.3} & Updating Job in real-time & Quantum software development is hampered by long job execution waiting times due to scarce resources and maintenance, compounded by the critical inability to modify circuits once queued, severely hindering agile progress.\\  
\end{tabular}
};
\draw [rounded corners=.5em] (table.north west) rectangle (table.south east);
\end{tikzpicture}
\label{tab:continuouschanges}
\end{table}

\begin{description}
\item[CC.1 Quantum SDK Control Version]
\mbox{ }
\\
The rapid advancement and fierce competition among quantum computer providers lead to frequent and often unannounced updates in their Software Development Kits (SDKs). These updates, driven by continuous innovation, frequently introduce significant improvements, new functionalities, and simultaneously, deprecated functions or altered APIs. While essential for pushing the boundaries of quantum computing, this constant evolution presents a substantial hurdle for software developers. The inherent instability can cause significant incompatibility errors in developed quantum software, particularly during critical phases such as transpilation, optimization, and execution on diverse quantum hardware back-ends. This dynamic environment poses a considerable challenge for developing stable, reliable, and maintainable quantum solutions that can endure over time without constant manual intervention.

To effectively address this critical challenge, it's imperative to develop sophisticated, automated tools and methodologies. These tools must be capable of intelligently identifying, analyzing, and automatically updating quantum software to the latest SDK versions. Crucially, such systems should incorporate robust automated testing frameworks designed to proactively detect any incompatibility issues or performance regressions within the production codebase immediately following an SDK update. If the automated verification process identifies errors, the system's output should be highly granular and actionable, clearly indicating the specific failures, their root causes, and their precise locations within the code, facilitating rapid debugging and remediation. Conversely, upon successful verification, the system should enable an automated, seamless deployment of the newly updated service. This automated deployment should strategically manage the replacement of existing instances running on older service versions with the newly validated ones, ensuring minimal disruption and continuous service availability. This proactive approach is vital for fostering agility and resilience in the face of rapidly evolving quantum hardware and software landscapes.
\\
\item[CC.2 Dynamic Resource Allocation]
\mbox{ }
\\
Dynamic resource allocation in quantum computing remains a major challenge due to the limited availability and highly specialized nature of quantum hardware. In contrast to classical computing, where resources are generally abundant and easily interchangeable, QPUs are scarce, expensive, and exhibit unique characteristics that influence algorithmic performance. This limitation, combined with long queue times and the heterogeneous capabilities of existing QPUs, makes efficient resource utilization essential but particularly difficult. Within this complex context, the concept of liquidity emerges as a key enabler. By giving both classical and quantum software components liquid-like properties, it becomes possible to achieve the flexibility required for applications to dynamically adapt and operate across the quantum-classical continuum. In this way, resources can be optimally leveraged at any given moment to ensure efficient execution and maximize throughput.

This challenge can be addressed through the development of adaptive quantum-classical orchestration layers that go beyond basic job scheduling. Such layers would analyze QPU availability, workload conditions, and specific job requirements in an intelligent manner. At the same time, the use of fine-grained quantum microservices is recommended, allowing complex algorithms to be divided into smaller tasks that can be executed on the most suitable QPU, or even on classical resources when immediate quantum access is unavailable. Applications should also be designed to support real-time shifting within the quantum-classical continuum, enabling them to dynamically reroute or approximate computations based on current resource conditions. In addition, proactive resource monitoring and prediction would help anticipate future QPU states, improving job placement decisions. Finally, automated quantum program transpilation and optimization for different hardware targets would ensure efficient execution regardless of the specific QPU selected.

By applying these strategies, quantum applications would no longer be constrained by fixed hardware or long waiting times. Instead, they would gain liquidity, seamlessly flowing across the quantum-classical continuum, adapting dynamically to resource availability, and accelerating the development and deployment of practical quantum solutions.
\\
\item[CC.3 Updating Jobs in Real-Time]
\mbox{ }
\\
Quantum software development grapples with a significant bottleneck, the protracted waiting periods for job execution. This delay is primarily driven by the high demand for scarce quantum computing resources from researchers across academia and industry, compounded by necessary maintenance windows for the quantum computers themselves. Consequently, it's not uncommon for a quantum job to experience execution delays of several days. While efforts are underway to maximize quantum machine utilization through techniques like simultaneous circuit execution \cite{Romero2024c}, an additional challenge arises: the inability to modify a quantum circuit once it has been submitted to the queue.

This lack of in-queue modifiability presents several critical issues. For instance, if an urgent software update becomes available, perhaps addressing a newly discovered bug or improving an algorithm's efficiency, developers are currently unable to apply these changes to jobs already awaiting execution. 
The inability to implement such dynamic changes without entirely resubmitting a job exacerbates the already long waiting times and hampers agile development and iterative refinement of quantum software. Addressing this limitation is crucial for improving developers' productivity and accelerating quantum research.

\end{description}
\subsection{Pricing configuration space}
\label{subsec:pricing_configurtaion_space}

The design and management of pricing strategies in SOQ systems present novel and underexplored challenges, distinct from those in classical cloud computing. Unlike traditional SaaS environments, where pricing can be based on well-understood metrics such as CPU hours, storage, or bandwidth, quantum computing introduces pricing variables tied to the probabilistic and hardware-dependent nature of quantum executions. These include factors such as the number of shots, qubit coherence time, gate fidelity, error rates, execution priority, etc.

This complexity gives rise to a \textbf{multidimensional pricing configuration space} in SOQ, where pricing is not only a business model issue, but a core technical challenge that affects orchestration, scheduling, and service-level guarantees. In Table \ref{tab:pricing}, we identify four main challenges in this context.

\begin{table}[!ht]
\caption{Pricing Configuration Space Challenges for SOQ.}
\centering
\begin{tikzpicture}
\node (table) [inner sep=0pt] {
\begin{tabular}{c|p{4.3cm}|p{8cm}}
    \rowcolor{blue!10}
    \textbf{\#} & \textbf{Challenge} & \textbf{Description} \\
    \hline
    \textbf{PCs.1} & Multidimensional and Hardware-Dependent Pricing Models. & Quantum pricing depends on heterogeneous hardware back-ends with different fidelities, queue times, and guarantees, requiring dynamic, provider-specific models that jointly optimize cost, quality, and availability. \\ 
    \hline
    \textbf{PCs.2} & Hybrid Pricing in Quantum-Classical Workflows. & Hybrid workflows combine asymmetric cost models (time-based for classical vs. shot-based for quantum), demanding unified pricing schemes that balance fidelity, cost, and performance. \\ 
    \hline
    \textbf{PCs.3} & Dynamic Access and Priority-Based Pricing. & Scarcity of quantum hardware drives priority-based and pay-for-priority tiers, introducing temporal volatility into costs and limiting flexibility once jobs are queued. \\  
    \hline
    \textbf{PCs.4} & Configuration Complexity and Economic Feasibility. & Complex pricing options create overwhelming configuration spaces, requiring automation and optimization tools to ensure economic feasibility.\\  
\end{tabular}
};
\draw [rounded corners=.5em] (table.north west) rectangle (table.south east);
\end{tikzpicture}
\label{tab:pricing}
\end{table}

\begin{description}
  
\item[PCs.1 Multidimensional and Hardware-Dependent Pricing Models]
\mbox{ }
\\
Quantum services are executed on heterogeneous hardware back-ends (QPUs), each with specific physical characteristics that influence both cost and performance. Providers like IBM, IonQ or Rigetti expose devices with varying fidelities, queue times, and execution constraints. As a result, a single quantum task may incur different costs depending on where and when it is executed. This creates the need for \textbf{pricing models that are dynamic, per-provider, and sensitive to quantum hardware properties}. Furthermore, advanced pricing tiers may offer guarantees on error thresholds or availability windows, further complicating the pricing space.
From a service binding perspective, this challenge directly intersects with the \textit{QoS-aware binding/composition problem} in service-oriented architectures \cite{Canfora2008}. Traditionally, binding a task to a provider involves evaluating a solution space defined by static QoS attributes. In SOQ with pricings, however, each provider introduces a set of pricing-dependent configurations, e.g., shot count limits, fidelity guarantees, or priority access, that effectively multiply the number of feasible bindings. Consequently, the solution space for the binding problem expands exponentially \cite{GarciaFernandez2024}, requiring new approaches to evaluate and rank execution plans that jointly optimize for cost, quality, and availability under pricing constraints.
\\
\item[PCs.2 Hybrid Pricing in Quantum-Classical Workflow]
\mbox{ }
\\
In SOQ, pricing must account for hybrid service compositions where quantum subroutines are embedded within larger classical workflows. While classical services are typically priced per compute hour or API call, quantum components are priced per shot or per circuit execution. The \textbf{asymmetric cost model} of hybrid services poses significant challenges for orchestrators and clients, who must balance fidelity, cost, and performance trade-offs in real time. Hybrid pricing models should allow combining different metrics (e.g., time-based for classical, shot-based for quantum) in a unified representation.
\\
\item[PCs.3 Dynamic Access and Priority-Based Pricing]
\mbox{ }
\\
Given the scarcity of quantum hardware, many providers implement \textbf{priority-based queueing mechanisms} or ``pay-for-priority'' tiers, where higher payment guarantees faster or more stable execution. This introduces temporal volatility into pricing. A task's final cost may depend on queue times, cancellations, or noise levels at the moment of execution. Moreover, current systems lack mechanisms for \textbf{real-time modification of jobs once queued}, limiting the ability to adapt to pricing changes after submission.
\\
\item[PCs.4 Configuration Complexity and Economic Feasibility]
\mbox{ }
\\
Users must navigate a large space of configurable pricing options,shots, fidelity constraints, noise-aware rerouting, execution priority, SLA guarantees, which can quickly lead to \textbf{overwhelming configuration spaces} \cite{GarciaFernandez2024}. These options must be analyzed not only for correctness \cite{GarciaFernandez2025} but for \textbf{economic feasibility}, especially in recurring or mission-critical scenarios. Techniques from classical pricing automation (e.g., iPricings or constraint optimization) may be adapted to quantum systems to support automated pricing validation, recommendation, and optimization. In the context of SOQ, quantum-specific properties such as qubit count, noise levels, and number of shots must be treated as declarative, contract-aware parameters within the pricing and service description layers. These parameters directly affect both the cost and feasibility of a service invocation and must therefore be seamlessly integrated into SLA definitions and exposed as part of the service metadata. Unlike QSOC, where such constraints are often addressed by introducing auxiliary validation tasks or imperatively altering the workflow, SOQ aspires to a declarative approach that avoids polluting the business logic. This means that service composition and selection mechanisms should validate whether the quantum backend can satisfy the resource and fidelity requirements of a given task at design or planning time, not during execution. By formalizing these parameters as part of the pricing configuration space, SOQ enables dynamic service negotiation, fallback mechanisms, and workflow optimization, while preserving architectural modularity and abstraction.

\end{description}
\subsection{Workforce training}
\label{subsec:workforce_traning}

As quantum computing advances, the lack of skilled professionals with expertise in quantum algorithms, hybrid quantum-classical systems, and SOC integration presents a significant barrier to adoption \cite{AparicioMorales2024}. Unlike classical computing, which has a well-established developer ecosystem with standardized tools and best practices, quantum computing is still in its early stages, requiring specialized knowledge. Training a workforce capable of developing, deploying, and maintaining quantum services is therefore critical to ensuring the success of the broader quantum software ecosystem \cite{Haghparast2024}.

Workforce training is crucial for interoperability, as developers must integrate quantum services with classical IT infrastructures. A lack of expertise in hybrid architectures can hinder quantum adoption. Similarly, platform independence depends on skilled professionals who can adapt applications across different hardware providers, avoiding vendor lock-in. Demand and capacity management also requires trained engineers to optimize workloads and schedule quantum-classical computations efficiently, preventing resource waste and high costs. Moreover, complexity management, continuous changes, and pricing models require expertise in quantum computing economics, hardware constraints, and evolving algorithms. Without ongoing training, scaling quantum applications beyond NISQ-era devices will be challenging. Investing in workforce development through education, industry collaboration, and standardized training is essential to ensure SOQ evolves, integrates into enterprises, and delivers real-world benefits.

According to these insights, in Table \ref{tab:training} we identify several relevant challenges in workforce training development.

\begin{table}[!ht]
\caption{Workforce Training Challenges for SOQ.}
\centering
\begin{tikzpicture}
\node (table) [inner sep=0pt] {
\begin{tabular}{c|p{4.3cm}|p{8cm}}
    \rowcolor{blue!10}
    \textbf{\#} & \textbf{Challenge} & \textbf{Description} \\
    \hline
    \textbf{WT.1} & Shortage of qualified professionals. & The current workforce lacks sufficient individuals with the specialized expertise required in quantum computing. \\ 
    \hline
    \textbf{WT.2} & Immaturity of the field and technical complexity. & Early stage of development of the field, with high technical complexity that hinders widespread understanding. \\  
    \hline
    \textbf{WT.3} & Inherently interdisciplinary nature of quantum development. & Effective work in quantum computing demands integrated knowledge across physics, computer science, engineering, and mathematics. \\  
    \hline
    \textbf{WT.4} & Insufficient or inadequate educational offerings. & Educational programs in quantum computing are limited in availability and scope.\\ 
\end{tabular}
};
\draw [rounded corners=.5em] (table.north west) rectangle (table.south east);
\end{tikzpicture}
\label{tab:training}
\end{table}

\begin{description}

\item[WT.1 Shortage of qualified professionals and a low critical mass of researchers]
\mbox{ }
\\
There is a significant scarcity of professionals with specialized knowledge in Quantum Technologies, particularly in the area of Quantum Computing. To address the shifts of this new paradigm, an increase in personnel with this expertise is required. Currently, there is still a low critical mass of researchers in QSE, which underscores the urgent need to train new researchers and students in the specific aspects of this field. Furthermore, the demand for qualified professionals in quantum software development is growing rapidly, creating an urgent need to fill job vacancies in the coming years.
\\
\item[WT.2 Immaturity of the field and technical complexity]
\mbox{ }
\\
Current quantum computers are noisy and susceptible to errors, which limits their practical utility for complex computations and necessitates sophisticated error correction techniques to mitigate these problems. Likewise, many of the elements of the infrastructure, tools, and techniques needed to develop quantum software solutions are in their early stages. This implies that training must not only address advanced theoretical concepts but also the practical limitations and challenges of current technology. The lack of standards also affects both technical progress and commercial adoption. This hinders interoperability between platforms, programming languages, and hardware, scalability, i.e., difficulty in expanding capabilities and efficiently connecting multiple systems, and collaborative innovation among scientific, academic, and industrial communities. The absence of standards in quantum computing creates unnecessary complexity for learners, discourages entry-level participation, and hinders the development of cohesive educational programs. As the field matures, establishing educational standards and cross-platform compatibility will be essential for building a broad, skilled quantum workforce.
Another key factor that limits the opportunities for learning quantum computing is the high cost of accessing quantum infrastructure. It significantly limits the democratization of knowledge and practical training opportunities in quantum systems education, while also amplifying inequalities between institutions worldwide. To make quantum education more inclusive and sustainable, it is essential to develop supported access models, shared affordable infrastructure, advanced free or open-source simulators, and promote strong partnerships between industry, government, and academia.
\\
\item[WT.3 Inherently interdisciplinary nature of quantum development]
\mbox{ }
\\
Quantum development requires fundamental contributions from mathematicians and physicists, demanding collaborative approaches for its advancement. Future professionals must possess a foundational theoretical understanding of quantum physics principles (fundamental concepts, mathematical frameworks, qubit dynamics) and the underlying physical principles of quantum technologies. Additionally, they need advanced expertise in a specific aspect of quantum technology and a keen understanding of the interrelationships among various facets of quantum technology and classical systems, including the integration of hybrid quantum-classical systems.
\\
\item[WT.4 Insufficient or inadequate educational offerings]
\mbox{ }
\\
Although the academic offerings in quantum computing are growing, with courses and master's degrees provided by various platforms (such as EdX or Udemy) and organizations (like IBM, Google, and Microsoft), most of these offerings focus on general introductory aspects or deal with quantum computing at a low level. There is a relative scarcity in the educational offerings that address the more specific and advanced aspects of quantum computing. For the implementation of this technology in productive sectors, it is crucial to increase the number of qualified individuals who can participate in research and development projects in the field.

Therefore, to build a professional profile, it is essential to understand the professional requirements established by experts and companies, such as those outlined in the European Competence Framework for Quantum Technologies (ECFQT). These requirements not only cover a solid theoretical foundation and advanced expertise in specific areas but also advanced technical skills such as understanding quantum computing principles, qubits, quantum gates, circuits, and algorithms (like Shor or Grover), quantum hardware technologies, quantum programming languages and SDKs, quantum error correction, quantum circuit optimization, and quantum software applications in real-world use cases. Furthermore, the need for practical experience through laboratories, simulations, and applied projects is emphasized, along with "soft skills".
 
\end{description}

In summary, the main challenge is to bridge the gap between the increasing demand for specialized talent in QSE and the limited supply of educational programs that provide the necessary depth and practical focus for such a complex and interdisciplinary field. Many initiatives, educational centers, and universities have detected this need and have begun to offer specialized courses and master's degrees in this field. A clear example is the role of the \textit{European initiative Quantum Flagship} \cite{Riedel2019}, which among its objectives, is the training and development of talent in the quantum field. Specifically, the initiative will promote various programs to educate and train the next generation of scientists, engineers, and professionals who will work in this emerging industry. DigiQ (Digitally Enhanced Quantum Technology Master) is the primary workforce development project of the Quantum Flagship. It is funded by a Euro 17.6 million grant over four years through the European Commission’s Digital Europe Programme. There are industrial associations specifically dedicated to promoting the use and adoption of quantum computing globally. These organizations bring together companies, universities, startups, and governments to collaborate on growing the quantum ecosystem. For instance, the \textit{European Quantum Industry Consortium} (QuIC). The aim of this association is to accelerate the industrial development of quantum technologies in Europe. They also work in collaboration with the Quantum Flagship initiative.


\section{From Adaptation to Native Integration: SOQ vs. QSOC}
\label{sec:soq}

This section highlights the main differences between SOQ and QSOC by analyzing the current technology stack for QSOC and proposing the new layers for SOQ.

\subsection{QSOC Technology Stack}

The modern computing landscape is the result of over six decades of technological evolution, refinement, and standardization. From the early days of transistor-based machines to today's globally distributed cloud systems, this layered stack of technologies forms the foundation upon which all software, ranging from web applications to mission-critical systems, has been built. The gradual stratification into layers has enabled modularity, portability, scalability, and increasingly abstracted interaction with underlying hardware.

At the lowest level, we find the hardware layer, which has evolved from vacuum tubes and early mainframes to powerful microprocessors, GPUs, and accelerators like FPGAs and TPUs. These components are responsible for the fundamental execution of machine-level instructions and are manufactured to support general-purpose, high-performance, or domain-specific computing.

On top of this sits the operating system (OS) layer, responsible for abstracting hardware complexity and providing interfaces for process management, memory allocation, I/O operations, and user-level execution. Operating systems like Linux, Windows, and macOS have become the dominant platforms in personal, server, and cloud computing environments, with specific variants for real-time or embedded systems.

With the rise of virtualization and lightweight computing, the container layer emerged, most notably through technologies like Docker. Containers offer isolated execution environments that are portable and fast to deploy, becoming central to microservices architectures. This was followed by container runtimes like containerd and CRI-O, which provide the core interfaces to manage container lifecycles.

To scale containerized applications, the orchestration layer became essential. Technologies such as Kubernetes allow for the management of thousands of containers across clusters of machines, offering high availability, load balancing, autoscaling, and fault tolerance. This layer transformed how distributed systems are deployed and maintained, introducing the “infrastructure as code” mindset.

Above this sits the cloud infrastructure layer, provided by global hyperscalers such as AWS, Microsoft Azure, and Google Cloud Platform. These platforms offer elastic computing, storage, and networking resources as services, allowing developers to focus on building applications without worrying about physical infrastructure. Concepts like pay-as-you-go billing, serverless computing, and global scalability are native to this layer.

On top of these foundational components, we find the service and application layer, which enables cloud computing, SOA, and microservices. Applications are decomposed into independent services that communicate through REST APIs, gRPC, or event-driven messaging systems. These services are developed using a range of programming languages and frameworks, such as Python, Java, Go or Node.js, and integrated into business workflows using orchestration tools like Apache Camel, Istio, or workflow engines.

At the same level as the cloud, we locate QSOC. While still rooted in classical infrastructure, QSOC represents the beginning of an architectural integration between classical service-oriented systems and quantum services. In QSOC, quantum resources are treated as remote services, invoked via classical workflows, without requiring the programmer to manage the complexity of quantum hardware directly. These services are often exposed through cloud APIs (e.g., AWS Braket or IBM Quantum) and are tightly coupled with classical layers for orchestration, preprocessing, and post-analysis.

This layered model, from hardware to QSOC, has enabled the modular, scalable, and resilient systems we rely on today. Each layer abstracts and builds upon the one beneath it, allowing innovation to flourish independently while maintaining interoperability and reliability.

\subsection{Proposed SOQ Layered Technology Stack}

The emerging field of quantum computing is progressing rapidly, but it remains in an early stage of development compared to the mature, multi-layered stack of classical computing, as we saw in the previous section. While classical systems continue to evolve at a rapid speed, quantum computing still relies on highly specialized hardware, fragmented software stacks, and vendor-specific execution environments.

With the SOQ paradigm, which we propose in this paper, we want to anticipate a future in which quantum computing can be abstracted, composed, and integrated as a native part of service-oriented architectures. However, for SOQ to become a reality, we believe it is necessary to develop a robust, layered quantum technology stack. This stack must provide not only access to quantum capabilities, but also interoperability, automation, pricing models, and service governance, similar to what classical stacks offer today. We understand that this transition must be gradual, but that the idea must be focused so that the entire industry and scientific community work along the same lines. For these reasons, we propose a layered model indicating existing quantum technology, the limitations that exist in each of the layers, and the requirements for achieving SOQ. To summarize, we provide the next Fig. \ref{fig:quantum_stack_summary} showing the proposed layers for SOQ, the current shortcomings, and the progress needed to achieve SOQ.

\begin{enumerate}
    \item \textbf{Quantum Hardware Layer}. At the base of the stack lies the quantum hardware layer, composed of diverse technologies such as superconducting qubits (IBM, Google), trapped ions (IonQ, Quantinuum), photonic qubits (Xanadu), and neutral atoms (Pasqal). Perhaps this layer is the most advanced and the one in which the greatest efforts and investments are being made. Although these technologies differ in coherence time, qubit connectivity, gate fidelity, and scalability. While they have enabled promising demonstrations of quantum advantage and commercial access, each vendor maintains proprietary hardware interfaces, performance, operability, and metrics, which limit interoperability and comparative benchmarking. To build a sustainable quantum ecosystem, the industry must move toward standardizing technology and developing more modular and interchangeable hardware interfaces, similar to what has been achieved with CPUs and GPUs in classical computing. 

    \item \textbf{Quantum Operating System and Runtime Layer}. Located at the top of the hardware stack, the control and runtime layer governs the low-level execution of quantum operations. There are already some offerings in this layer that include vendor-specific firmware and early runtime environments such as Qiskit Runtime (IBM), Cirq Runtime (Google), Amazon Braket (AWS), or FireOpal (Q-CTRL), which handle pulse-level control, job submission, and error mitigation. However, the quantum field still lacks a general-purpose operating system capable of scheduling processes, isolating workloads, or virtualizing resources. Moreover, unlike the classical domain where POSIX has provided a widely adopted standard for interoperability and abstraction, there is currently no equivalent framework guiding the design of quantum operating systems. This absence of a unifying standard hinders the development of robust multi-user environments and complicates integration with high-level orchestration tools. Future quantum systems will require shared runtimes, portable low-level APIs, and unified control layers that can manage execution across diverse devices in a predictable and scalable manner.

    \begin{figure}[H]
    \centering
       \caption{Summary of Quantum Technology Stack: Deficiencies and Research Priorities.}
        \includegraphics[width=0.9\textwidth]{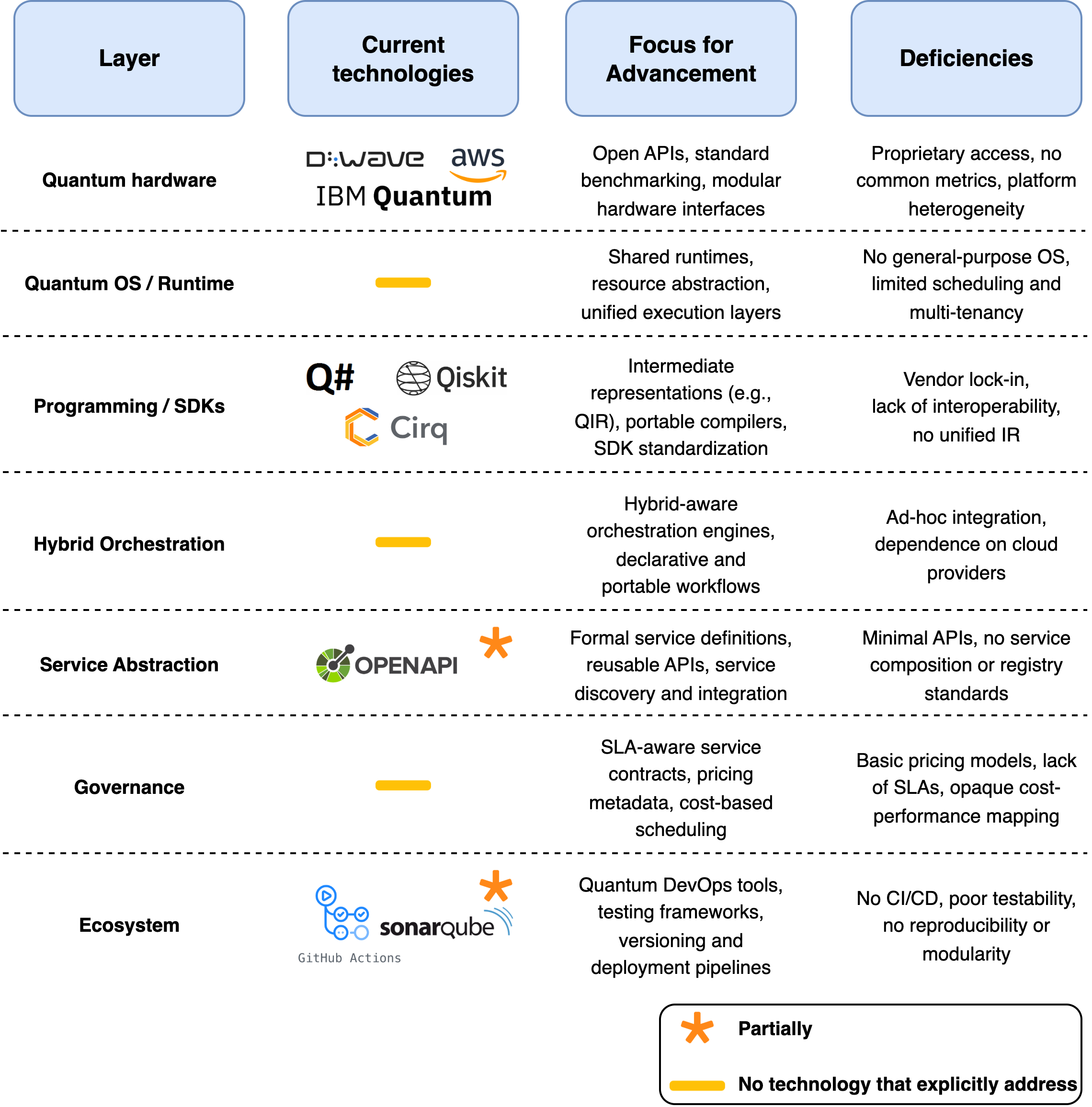}
        \label{fig:quantum_stack_summary}
    \end{figure}
    
    \item \textbf{Programming and SDK Layer}. The programming layer allows developers to write and test quantum applications using high-level languages and SDKs. Popular frameworks such as Qiskit, Cirq, PennyLane, and Q\# have democratized quantum programming by enabling circuit design, hybrid algorithm development, and integration with classical logic. Despite this progress, most SDKs are tightly coupled to their underlying (classical) platforms, limiting cross-vendor compatibility and confining developers to specific ecosystems. Furthermore, the absence of a universally adopted intermediate representation hinders code portability and toolchain interoperability. Research and development should focus on defining common IRs (e.g., QIR), improving compiler portability, and creating toolchains that allow quantum programs to run on different hardware backends without rewriting or vendor-specific tuning. These efforts align with the emerging field of Quantum Software Engineering, which emphasizes proper abstraction mechanisms, enhanced expressiveness of quantum programs, and greater developer productivity, thereby complementing the technical advances in interoperability and portability \cite{Piattini2020}.

    \item \textbf{Hybrid Classical-Quantum Orchestration Layer}. Since current quantum devices are still in the NISQ era, most useful algorithms are based on hybrid execution models. Platforms such as Amazon Braket and IBM provide hybrid task orchestration that allows classical systems to coordinate the execution of quantum circuits. However, these orchestration mechanisms are often ad hoc, non-standardized, and dependent on the host cloud provider. There is little support for dynamic task allocation, distributed orchestration, or workflow portability. To enable scalable hybrid applications, the ecosystem must develop orchestration frameworks that are aware of quantum constraints such as noise, coherence time, and queue latency, and that are capable of intelligently allocating tasks at execution time between classical and quantum resources using flexible, declarative workflows. A key challenge is also to determine when quantum execution is truly advantageous compared to classical alternatives. Such decisions depend not only on runtime factors (e.g., noise, coherence, invocation overheads) but also on governance and pricing models that influence cost-effectiveness and sustainability.

    \item \textbf{Service Abstraction Layer}. A fundamental requirement of SOQ is the ability to expose quantum functionalities as services, with well-defined interfaces, standardized inputs and outputs, and service quality descriptors. Current platforms, such as IBM Quantum and AWS Braket, offer basic service interfaces through REST APIs or SDKs, but these are often limited to job submission and lack deeper composability features. Furthermore, there are no industry-wide conventions for describing quantum services, publishing them in registries, or integrating them into multiservice architectures. Advancing this layer will require formal service definition standards, composable APIs, service registries, and mechanisms for interface negotiation and semantic interoperability between heterogeneous quantum providers.

    \item \textbf{Governance}. Unlike classical computing, where SLAs, pricing models, and quota systems are well established, quantum computing lacks unified approaches to economic governance. Current pricing schemes are basic and tied to metrics such as the number of shots, qubit usage, or execution priority. There is minimal transparency regarding the relationship between cost and performance or service guarantees, and prices are often not displayed as part of the service interface. To launch large-scale quantum services, platforms must implement pricing metadata, SLA descriptors, and brokerage mechanisms that allow applications to select providers based on fidelity, execution time, and cost constraints. This also opens the door to cost-conscious orchestration and market-based resource optimization.

    \item \textbf{Ecosystem Layer}. In classical software engineering, DevOps practices such as version control, CI/CD processes, automated testing, and observability are the norm. In quantum software, these practices are virtually absent. Development is often done using notebooks and scripts, with limited tools for testing quantum behavior, simulating realistic noise, or tracking quantum job history. Furthermore, there is no support for reproducibility, dependency management, or modular code reuse. For SOQ to be viable, the ecosystem must provide DevOps pipelines integrated into SDKs, domain-specific testing frameworks, and versioning and packaging tools compatible with quantum technology. Beyond adapting classical DevOps practices, the ecosystem will likely need inherently quantum-oriented tools, such as dashboards or monitoring frameworks that expose quantum-specific execution properties (e.g., qubit usage against resource limits, noise levels, or average shots per program). These native observability capabilities would complement traditional pipelines and provide developers with meaningful feedback about workload efficiency and performance. This combination will enable professional-grade software engineering practices in quantum development and lower adoption barriers for enterprises and open-source contributors.
    
\end{enumerate}

\subsection{SOQ vs. QSOC}

The comparison between QSOC and the SOQ, proposed in this manuscript, reveals a fundamental change in how quantum capabilities are conceptualized, integrated, and composed within software architectures. While QSOC was an important first step in adapting classical service-oriented computing principles to support quantum operations, SOQ redefines the architectural foundations by making quantum services native elements of the system design, rather than mere extensions of classical infrastructures.

\begin{figure}[ht]
\centering
   \caption{QSOC vs. SOQ: Key Differences.}
    \includegraphics[width=0.8\textwidth]{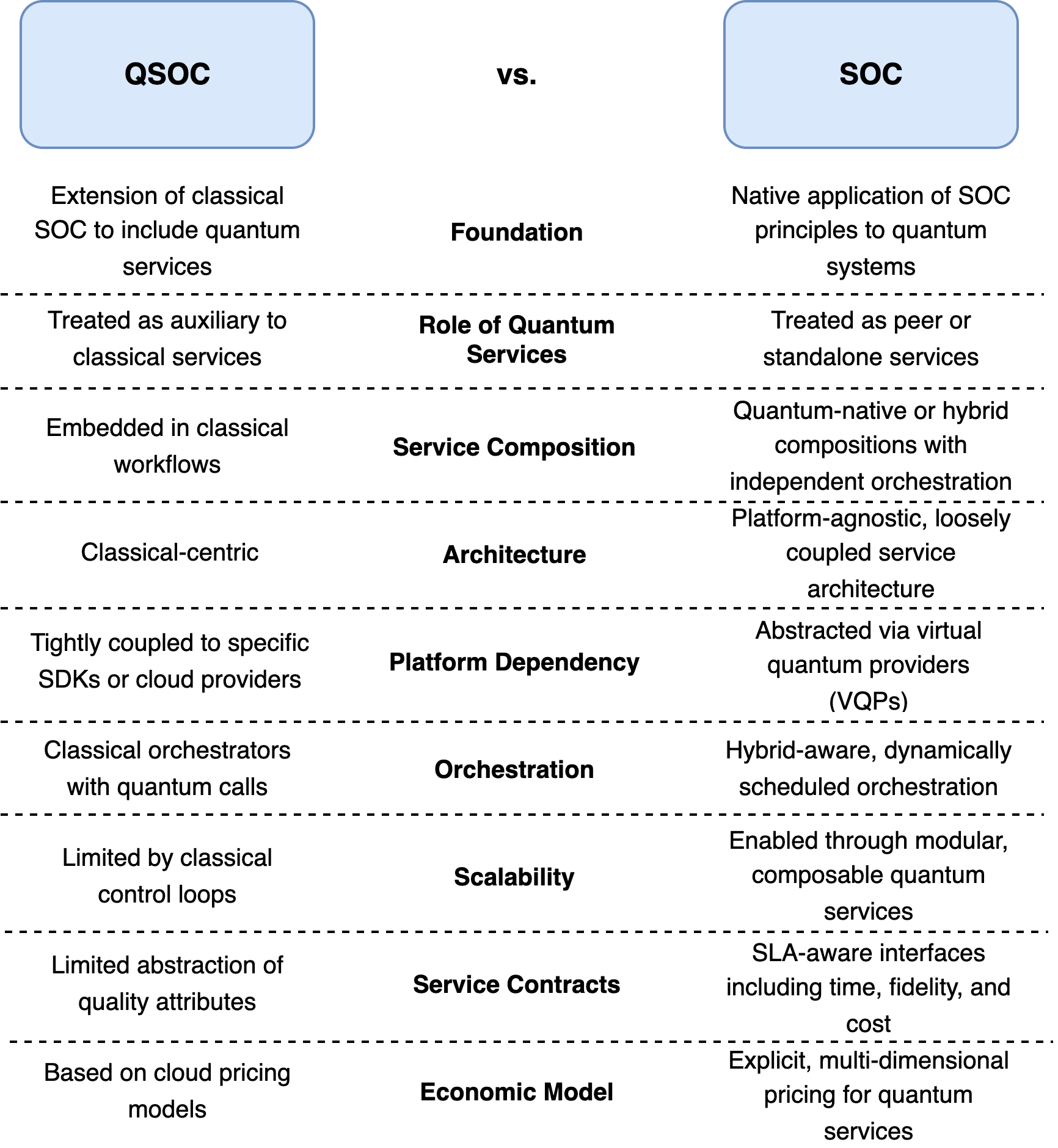}
    \label{fig:soq-vs-qsoc}
\end{figure}

In a nutshell, and as can be seen in Fig. \ref{fig:soq-vs-qsoc}, we believe that these differences between the two paradigms represent a qualitative jump toward treating quantum software as a fundamental element in service-oriented computing. The layered architecture, modular abstractions, orchestration models, and pricing mechanisms introduced by SOQ are key factors for the future of large-scale quantum software engineering. 

In QSOC, the architecture is predominantly classical-centric. Quantum components are typically invoked as specialized, back-end services that operate under classical orchestration and control. This model assumes that the quantum functionality is subordinate or complementary to a larger classical workflow. As a result, service composition, orchestration, and deployment remain bound by the constraints and assumptions of classical software engineering.

In contrast, SOQ advocates a new paradigm where quantum services are autonomous, interoperable, and composable at the same level of abstraction as classical services. Rather than embedding quantum operations within classical applications, SOQ would allow the development of quantum-native service compositions, including pure quantum workflows, hybrid orchestration, and platform-independent service ecosystems. This design would promote dynamic and adaptable quantum software interactions, reusability, and discoverability, which will be essential for scalability and long-term maintainability in quantum software systems.

Another major distinction lies in orchestration and execution control. QSOC relies on classical orchestrators, which may not be optimized to manage quantum-specific concerns. The actual challenge emerges in the interplay between traditional QoS aspects, typically central to orchestration and binding decisions, and the integration of quantum properties, such as noise sensitivity, probabilistic behavior, execution queue latency, or fidelity guarantees, as part of the QoS of the system. This interplay gives rise to new challenges in ensuring efficient and reliable orchestration in hybrid quantum-classical environments.

This also significantly influences economic models for accessing quantum computing platforms. In QSOC, prices tend to reflect classic cloud-based services, focusing on resource time and performance, although other parameters such as number of qubits, tasks, and/or shots are also included. In contrast, SOQ would allow for the explicit integration of multidimensional pricing models that take into account factors such as fidelity targets, priority execution, or the calibration of the quantum machines themselves, enabling cost-conscious orchestration strategies and more transparent resource governance.

Finally, in terms of platform abstraction, QSOC architectures are typically tied to specific SDKs or quantum back-ends, limiting portability. SOQ introduces layers such as Virtual Quantum Providers (VQPs) and platform-agnostic APIs that support seamless integration across heterogeneous quantum hardware and simulators, paving the way for ecosystem-wide interoperability.

\section{Conclusion \& Future perspectives}
\label{sec:conclusion}

This work has introduced and articulated Service-Oriented Quantum (SOQ) as a forward-looking paradigm for structuring quantum software systems using the principles of service orientation. Building upon the limitations identified in the Quantum Service-Oriented Computing (QSOC) model, SOQ emphasizes the importance of treating quantum services as native, first-class components, autonomous, composable, and platform-agnostic. By extending the lessons learned from classical service architectures, SOQ enables modular design, hybrid orchestration, and interoperable deployments, while accounting for the unique characteristics of quantum hardware and software.

To support this vision, we presented a layered SOQ technology stack, outlining the necessary components and abstractions ranging from hardware to governance and DevOps. We identified key challenges in areas such as interoperability, hybridity, pricing configuration, circuit modularity, and workforce training, and discussed how addressing these challenges will be critical for the evolution of quantum software engineering. Through case studies and references to recent research, we demonstrated that momentum is already building toward many of these goals, yet much work remains.

Looking ahead, the future of SOQ lies in the co-evolution of quantum and classical infrastructures, and in the creation of a robust ecosystem of services, platforms, and engineering tools. As the industry transitions from NISQ devices toward more scalable and error-resilient quantum hardware, SOQ can serve as the architectural backbone for building sustainable, reusable, and verifiable quantum applications. In parallel, we foresee the emergence of cross-provider orchestration engines, declarative service composition languages for quantum workflows, and AI-driven design and testing tools tailored to the quantum domain.

Ultimately, SOQ provides a practical and extensible framework for integrating quantum computing into the broader landscape of software engineering. It opens new avenues for research in distributed systems, architecture design, economic models, and development methodologies in the quantum era. By grounding quantum software in proven engineering principles while embracing its unique constraints and opportunities, SOQ charts a path toward a scalable, service-driven quantum future.

\begin{acks}
    This work has been partially funded by the European Union “Next GenerationEU /PRTR”, by the Ministry of Science, Innovation and Universities (TED2021-130913B-I00, and PDC2022-133465-I00). It is also supported by QSERV project (PID2021-1240454OB-C31), PERSEO project (PID2021-126227NB-C21), and ATHENA project (PID2024-155693NB-C41) funded by the Spanish Ministry of Science and Innovation and ERDF; by European Union under the Agreement — 101083667 of the Project “TECH4E -Tech4effiencyEDlH” Call: DIGITAL-2021-EDlH-01 supported by the European Commission through the Digital Europe Program; by the Regional Ministry of Economy, Science and Digital Agenda of the Regional Government of Extremadura (GR24099). Supported by grant PRE2022-102070, financed by  MCIN/AEI/10.13039/501100011033, ``FEDER/EU'', and FSE+.
\end{acks}

\bibliographystyle{ACM-Reference-Format}
\bibliography{references}


\begin{thebibliography}{88}


\ifx \showCODEN    \undefined \def \showCODEN     #1{\unskip}     \fi
\ifx \showISBNx    \undefined \def \showISBNx     #1{\unskip}     \fi
\ifx \showISBNxiii \undefined \def \showISBNxiii  #1{\unskip}     \fi
\ifx \showISSN     \undefined \def \showISSN      #1{\unskip}     \fi
\ifx \showLCCN     \undefined \def \showLCCN      #1{\unskip}     \fi
\ifx \shownote     \undefined \def \shownote      #1{#1}          \fi
\ifx \showarticletitle \undefined \def \showarticletitle #1{#1}   \fi
\ifx \showURL      \undefined \def \showURL       {\relax}        \fi
\providecommand\bibfield[2]{#2}
\providecommand\bibinfo[2]{#2}
\providecommand\natexlab[1]{#1}
\providecommand\showeprint[2][]{arXiv:#2}

\bibitem[Aaronson(2010)]%
        {Aaronson2010}
\bibfield{author}{\bibinfo{person}{Scott Aaronson}.} \bibinfo{year}{2010}\natexlab{}.
\newblock \showarticletitle{BQP and the polynomial hierarchy}. In \bibinfo{booktitle}{\emph{Proceedings of the Forty-Second ACM Symposium on Theory of Computing}} (Cambridge, Massachusetts, USA) \emph{(\bibinfo{series}{STOC '10})}. \bibinfo{publisher}{Association for Computing Machinery}, \bibinfo{address}{New York, NY, USA}, \bibinfo{pages}{141–--150}.
\newblock
\showISBNx{9781450300506}
\href{https://doi.org/10.1145/1806689.1806711}{doi:\nolinkurl{10.1145/1806689.1806711}}


\bibitem[Ahmad et~al\mbox{.}(2024)]%
        {Ahmad2024}
\bibfield{author}{\bibinfo{person}{Aakash Ahmad}, \bibinfo{person}{Ahmed~B Altamimi}, {and} \bibinfo{person}{Jamal Aqib}.} \bibinfo{year}{2024}\natexlab{}.
\newblock \showarticletitle{A reference architecture for quantum computing as a service}.
\newblock \bibinfo{journal}{\emph{Journal of King Saud University-Computer and Information Sciences}} \bibinfo{volume}{36}, \bibinfo{number}{6} (\bibinfo{year}{2024}), \bibinfo{pages}{1--18}.
\newblock
\href{https://doi.org/10.1016/j.jksuci.2024.102094}{doi:\nolinkurl{10.1016/j.jksuci.2024.102094}}


\bibitem[Ali and Yue(2023)]%
        {Ali2023}
\bibfield{author}{\bibinfo{person}{Shaukat Ali} {and} \bibinfo{person}{Tao Yue}.} \bibinfo{year}{2023}\natexlab{}.
\newblock \showarticletitle{On the Need of Quantum-Oriented Paradigm}. In \bibinfo{booktitle}{\emph{Proceedings of the 2nd International Workshop on Quantum Programming for Software Engineering}} (San Francisco, CA, USA) \emph{(\bibinfo{series}{QP4SE 2023})}. \bibinfo{publisher}{Association for Computing Machinery}, \bibinfo{address}{New York, NY, USA}, \bibinfo{pages}{17–--20}.
\newblock
\showISBNx{9798400703768}
\href{https://doi.org/10.1145/3617570.3617868}{doi:\nolinkurl{10.1145/3617570.3617868}}


\bibitem[Alvarado-Valiente et~al\mbox{.}(2023a)]%
        {alvarado2023quantum}
\bibfield{author}{\bibinfo{person}{Jaime Alvarado-Valiente}, \bibinfo{person}{Javier Romero-{\'A}lvarez}, \bibinfo{person}{Ana D{\'\i}az}, \bibinfo{person}{Mois{\'e}s Rodr{\'\i}guez}, \bibinfo{person}{Ignacio Garc{\'\i}a-Rodr{\'\i}guez}, \bibinfo{person}{Enrique Moguel}, \bibinfo{person}{Jose Garcia-Alonso}, {and} \bibinfo{person}{Juan~M Murillo}.} \bibinfo{year}{2023}\natexlab{a}.
\newblock \showarticletitle{Quantum services generation and deployment process: a quality-oriented approach}. In \bibinfo{booktitle}{\emph{International Conference on the Quality of Information and Communications Technology}}. Springer, \bibinfo{pages}{200--214}.
\newblock
\href{https://doi.org/10.1007/978-3-031-43703-8_15}{doi:\nolinkurl{10.1007/978-3-031-43703-8_15}}


\bibitem[Alvarado-Valiente et~al\mbox{.}(2022)]%
        {Alvarado2022}
\bibfield{author}{\bibinfo{person}{Jaime Alvarado-Valiente}, \bibinfo{person}{Javier Romero-{\'A}lvarez}, \bibinfo{person}{Jose Garcia-Alonso}, {and} \bibinfo{person}{Juan~M Murillo}.} \bibinfo{year}{2022}\natexlab{}.
\newblock \showarticletitle{A guide for quantum web services deployment}. In \bibinfo{booktitle}{\emph{International Conference on Web Engineering}}. \bibinfo{publisher}{Springer}, \bibinfo{address}{Cham}, \bibinfo{pages}{493--496}.
\newblock
\href{https://doi.org/10.1007/978-3-031-09917-5_42}{doi:\nolinkurl{10.1007/978-3-031-09917-5_42}}


\bibitem[Alvarado-Valiente et~al\mbox{.}(2023b)]%
        {alvarado2023devops}
\bibfield{author}{\bibinfo{person}{Jaime Alvarado-Valiente}, \bibinfo{person}{Javier Romero-{\'A}lvarez}, \bibinfo{person}{Enrique Moguel}, {and} \bibinfo{person}{Jos{\'e} Garc{\'\i}a-Alonso}.} \bibinfo{year}{2023}\natexlab{b}.
\newblock \showarticletitle{Quantum web services orchestration and management using DevOps techniques}. In \bibinfo{booktitle}{\emph{International Conference on Web Engineering}}. Springer, \bibinfo{pages}{389--394}.
\newblock
\href{https://doi.org/10.1007/978-3-031-34444-2_33}{doi:\nolinkurl{10.1007/978-3-031-34444-2_33}}


\bibitem[Alvarado-Valiente et~al\mbox{.}(2024a)]%
        {alvarado2024orchestration}
\bibfield{author}{\bibinfo{person}{Jaime Alvarado-Valiente}, \bibinfo{person}{Javier Romero-{\'A}lvarez}, \bibinfo{person}{Enrique Moguel}, \bibinfo{person}{Jose Garc{\'\i}a-Alonso}, {and} \bibinfo{person}{Juan~M Murillo}.} \bibinfo{year}{2024}\natexlab{a}.
\newblock \showarticletitle{Orchestration for quantum services: The power of load balancing across multiple service providers}.
\newblock \bibinfo{journal}{\emph{Science of Computer Programming}}  \bibinfo{volume}{237} (\bibinfo{year}{2024}), \bibinfo{pages}{103139}.
\newblock
\href{https://doi.org/10.1016/j.scico.2024.103139}{doi:\nolinkurl{10.1016/j.scico.2024.103139}}


\bibitem[Alvarado-Valiente et~al\mbox{.}(2024b)]%
        {AlvaradoValiente2024}
\bibfield{author}{\bibinfo{person}{Jaime Alvarado-Valiente}, \bibinfo{person}{Javier Romero-{\'A}lvarez}, \bibinfo{person}{Enrique Moguel}, \bibinfo{person}{Jos{\'e} Garc{\'\i}a-Alonso}, {and} \bibinfo{person}{Juan~M. Murillo}.} \bibinfo{year}{2024}\natexlab{b}.
\newblock \showarticletitle{Technological diversity of quantum computing providers: a comparative study and a proposal for API Gateway integration}.
\newblock \bibinfo{journal}{\emph{Software Quality Journal}} \bibinfo{volume}{32}, \bibinfo{number}{1} (\bibinfo{year}{2024}), \bibinfo{pages}{53--73}.
\newblock
\showISBNx{1573-1367}
\href{https://doi.org/10.1007/s11219-023-09633-5}{doi:\nolinkurl{10.1007/s11219-023-09633-5}}


\bibitem[Aparicio-Morales et~al\mbox{.}(2024)]%
        {aparicio2024overview}
\bibfield{author}{\bibinfo{person}{{\'A}lvaro~M Aparicio-Morales}, \bibinfo{person}{Enrique Moguel}, \bibinfo{person}{Luis~Mariano Bibbo}, \bibinfo{person}{Alejandro Fernandez}, \bibinfo{person}{Jose Garcia-Alonso}, {and} \bibinfo{person}{Juan~M Murillo}.} \bibinfo{year}{2024}\natexlab{}.
\newblock \showarticletitle{An overview of quantum software engineering in Latin America}.
\newblock \bibinfo{journal}{\emph{Quantum Information Processing}} \bibinfo{volume}{23}, \bibinfo{number}{11} (\bibinfo{year}{2024}), \bibinfo{pages}{380}.
\newblock
\href{https://doi.org/10.1007/s11128-024-04586-5}{doi:\nolinkurl{10.1007/s11128-024-04586-5}}


\bibitem[Bharti et~al\mbox{.}(2022)]%
        {Bharti2022}
\bibfield{author}{\bibinfo{person}{Kishor Bharti}, \bibinfo{person}{Alba Cervera-Lierta}, \bibinfo{person}{Thi~Ha Kyaw}, \bibinfo{person}{Tobias Haug}, \bibinfo{person}{Sumner Alperin-Lea}, \bibinfo{person}{Abhinav Anand}, \bibinfo{person}{Matthias Degroote}, \bibinfo{person}{Hermanni Heimonen}, \bibinfo{person}{Jakob~S. Kottmann}, \bibinfo{person}{Tim Menke}, \bibinfo{person}{Wai-Keong Mok}, \bibinfo{person}{Sukin Sim}, \bibinfo{person}{Leong-Chuan Kwek}, {and} \bibinfo{person}{Al\'an Aspuru-Guzik}.} \bibinfo{year}{2022}\natexlab{}.
\newblock \showarticletitle{Noisy intermediate-scale quantum algorithms}.
\newblock \bibinfo{journal}{\emph{Rev. Mod. Phys.}}  \bibinfo{volume}{94} (\bibinfo{year}{2022}), \bibinfo{pages}{1--69}.
\newblock
Issue 1.
\href{https://doi.org/10.1103/RevModPhys.94.015004}{doi:\nolinkurl{10.1103/RevModPhys.94.015004}}


\bibitem[Binz et~al\mbox{.}(2013)]%
        {Binz2013}
\bibfield{author}{\bibinfo{person}{Tobias Binz}, \bibinfo{person}{Uwe Breitenb{\"u}cher}, \bibinfo{person}{Oliver Kopp}, {and} \bibinfo{person}{Frank Leymann}.} \bibinfo{year}{2013}\natexlab{}.
\newblock \showarticletitle{TOSCA: portable automated deployment and management of cloud applications}.
\newblock In \bibinfo{booktitle}{\emph{Advanced Web Services}}. \bibinfo{publisher}{Springer}, \bibinfo{address}{Cham}, \bibinfo{pages}{527--549}.
\newblock
\href{https://doi.org/10.1007/978-1-4614-7535-4_22}{doi:\nolinkurl{10.1007/978-1-4614-7535-4_22}}


\bibitem[Bisicchia et~al\mbox{.}(2025)]%
        {Bisicchia2025}
\bibfield{author}{\bibinfo{person}{Giuseppe Bisicchia}, \bibinfo{person}{Alessandro Bocci}, {and} \bibinfo{person}{Antonio Brogi}.} \bibinfo{year}{2025}\natexlab{}.
\newblock \showarticletitle{Quantum Executor: A Unified Interface for Quantum Computing}.
\newblock \bibinfo{journal}{\emph{arXiv preprint arXiv:2507.07597}} \bibinfo{volume}{1}, \bibinfo{number}{1} (\bibinfo{year}{2025}), \bibinfo{pages}{1--11}.
\newblock
\href{https://doi.org/10.48550/arXiv.2507.07597}{doi:\nolinkurl{10.48550/arXiv.2507.07597}}


\bibitem[Bisicchia et~al\mbox{.}(2024a)]%
        {Bisicchia2024c}
\bibfield{author}{\bibinfo{person}{Giuseppe Bisicchia}, \bibinfo{person}{Alessandro Bocci}, \bibinfo{person}{Jos{\'e} Garc{\'\i}a-Alonso}, \bibinfo{person}{Juan~M Murillo}, {and} \bibinfo{person}{Antonio Brogi}.} \bibinfo{year}{2024}\natexlab{a}.
\newblock \showarticletitle{Cut\&Shoot: Cutting \& Distributing Quantum Circuits Across Multiple NISQ Computers}. In \bibinfo{booktitle}{\emph{2024 IEEE International Conference on Quantum Computing and Engineering (QCE)}}, Vol.~\bibinfo{volume}{2}. \bibinfo{publisher}{IEEE}, \bibinfo{address}{Montreal, QC, Canada}, \bibinfo{pages}{187--192}.
\newblock
\href{https://doi.org/10.1109/QCE60285.2024.10276}{doi:\nolinkurl{10.1109/QCE60285.2024.10276}}


\bibitem[Bisicchia et~al\mbox{.}(2024b)]%
        {Bisicchia2024b}
\bibfield{author}{\bibinfo{person}{Giuseppe Bisicchia}, \bibinfo{person}{Giuseppe Clemente}, \bibinfo{person}{Jose Garcia-Alonso}, \bibinfo{person}{Juan Manuel~Murillo Rodr{\'\i}guez}, \bibinfo{person}{Massimo D'Elia}, {and} \bibinfo{person}{Antonio Brogi}.} \bibinfo{year}{2024}\natexlab{b}.
\newblock \showarticletitle{Distributing Quantum Computations, Shot-wise}.
\newblock \bibinfo{journal}{\emph{arXiv preprint arXiv:2411.16530}} \bibinfo{volume}{1}, \bibinfo{number}{1} (\bibinfo{year}{2024}), \bibinfo{pages}{1--22}.
\newblock
\href{https://doi.org/10.48550/arXiv.2411.16530}{doi:\nolinkurl{10.48550/arXiv.2411.16530}}


\bibitem[Bisicchia et~al\mbox{.}(2023a)]%
        {Bisicchia2023b}
\bibfield{author}{\bibinfo{person}{Giuseppe Bisicchia}, \bibinfo{person}{Jos{\'e} Garc{\'\i}a-Alonso}, \bibinfo{person}{Juan~M Murillo}, {and} \bibinfo{person}{Antonio Brogi}.} \bibinfo{year}{2023}\natexlab{a}.
\newblock \showarticletitle{Dispatching shots among multiple quantum computers: An architectural proposal}. In \bibinfo{booktitle}{\emph{2023 IEEE International Conference on Quantum Computing and Engineering (QCE)}}, Vol.~\bibinfo{volume}{2}. \bibinfo{publisher}{IEEE}, \bibinfo{address}{Bellevue, WA, USA}, \bibinfo{pages}{195--198}.
\newblock
\href{https://doi.org/10.1109/QCE57702.2023.10210}{doi:\nolinkurl{10.1109/QCE57702.2023.10210}}


\bibitem[Bisicchia et~al\mbox{.}(2023b)]%
        {Bisicchia2023}
\bibfield{author}{\bibinfo{person}{Giuseppe Bisicchia}, \bibinfo{person}{Jose Garc{\'\i}a-Alonso}, \bibinfo{person}{Juan~M Murillo}, {and} \bibinfo{person}{Antonio Brogi}.} \bibinfo{year}{2023}\natexlab{b}.
\newblock \showarticletitle{Distributing quantum computations, by shots}. In \bibinfo{booktitle}{\emph{International Conference on Service-Oriented Computing}}. \bibinfo{publisher}{Springer}, \bibinfo{address}{Cham}, \bibinfo{pages}{363--377}.
\newblock
\href{https://doi.org/10.1007/978-3-031-48421-6_25}{doi:\nolinkurl{10.1007/978-3-031-48421-6_25}}


\bibitem[Bisicchia et~al\mbox{.}(2024c)]%
        {Bisicchia2024}
\bibfield{author}{\bibinfo{person}{Giuseppe Bisicchia}, \bibinfo{person}{Jose García-Alonso}, \bibinfo{person}{Juan~M. Murillo}, {and} \bibinfo{person}{Antonio Brogi}.} \bibinfo{year}{2024}\natexlab{c}.
\newblock \showarticletitle{From Quantum Software Handcrafting to Quantum Software Engineering}. In \bibinfo{booktitle}{\emph{2024 IEEE International Conference on Software Analysis, Evolution and Reengineering - Companion (SANER-C)}}. \bibinfo{publisher}{IEEE}, \bibinfo{address}{Rovaniemi, Finland}, \bibinfo{pages}{149--150}.
\newblock
\href{https://doi.org/10.1109/SANER-C62648.2024.00026}{doi:\nolinkurl{10.1109/SANER-C62648.2024.00026}}


\bibitem[Canfora et~al\mbox{.}(2008)]%
        {Canfora2008}
\bibfield{author}{\bibinfo{person}{Gerardo Canfora}, \bibinfo{person}{Massimiliano Di~Penta}, \bibinfo{person}{Raffaele Esposito}, {and} \bibinfo{person}{Maria~Luisa Villani}.} \bibinfo{year}{2008}\natexlab{}.
\newblock \showarticletitle{A framework for QoS-aware binding and re-binding of composite web services}.
\newblock \bibinfo{journal}{\emph{Journal of Systems and Software}} \bibinfo{volume}{81}, \bibinfo{number}{10} (\bibinfo{year}{2008}), \bibinfo{pages}{1754--1769}.
\newblock


\bibitem[Chuang et~al\mbox{.}(1998)]%
        {Isaac1998}
\bibfield{author}{\bibinfo{person}{Isaac~L. Chuang}, \bibinfo{person}{Neil Gershenfeld}, {and} \bibinfo{person}{Mark Kubinec}.} \bibinfo{year}{1998}\natexlab{}.
\newblock \showarticletitle{Experimental Implementation of Fast Quantum Searching}.
\newblock \bibinfo{journal}{\emph{Phys. Rev. Lett.}}  \bibinfo{volume}{80} (\bibinfo{year}{1998}), \bibinfo{pages}{3408--3411}.
\newblock
Issue 15.
\href{https://doi.org/10.1103/PhysRevLett.80.3408}{doi:\nolinkurl{10.1103/PhysRevLett.80.3408}}


\bibitem[Clark and Stepney(2002)]%
        {Clark2002}
\bibfield{author}{\bibinfo{person}{John Clark} {and} \bibinfo{person}{Susan Stepney}.} \bibinfo{year}{2002}\natexlab{}.
\newblock \bibinfo{booktitle}{\emph{Proposed "Grand Challenge for Computing Research" Quantum Software Engineering}}.
\newblock \bibinfo{type}{{T}echnical {R}eport}. \bibinfo{institution}{University of York}.
\newblock
\urldef\tempurl%
\url{https://www.cs.york.ac.uk/nature/gc7/journeys.pdf}
\showURL{%
\tempurl}


\bibitem[Cranganore et~al\mbox{.}(2024)]%
        {Cranganore2024}
\bibfield{author}{\bibinfo{person}{Sandeep~Suresh Cranganore}, \bibinfo{person}{Vincenzo De~Maio}, \bibinfo{person}{Ivona Brandic}, {and} \bibinfo{person}{Ewa Deelman}.} \bibinfo{year}{2024}\natexlab{}.
\newblock \showarticletitle{Paving the way to hybrid quantum--classical scientific workflows}.
\newblock \bibinfo{journal}{\emph{Future Generation Computer Systems}}  \bibinfo{volume}{158} (\bibinfo{year}{2024}), \bibinfo{pages}{346--366}.
\newblock
\href{https://doi.org/10.1016/j.future.2024.04.030}{doi:\nolinkurl{10.1016/j.future.2024.04.030}}


\bibitem[Cross et~al\mbox{.}(2022)]%
        {Cross2022}
\bibfield{author}{\bibinfo{person}{Andrew Cross}, \bibinfo{person}{Ali Javadi-Abhari}, \bibinfo{person}{Thomas Alexander}, \bibinfo{person}{Niel De~Beaudrap}, \bibinfo{person}{Lev~S Bishop}, \bibinfo{person}{Steven Heidel}, \bibinfo{person}{Colm~A Ryan}, \bibinfo{person}{Prasahnt Sivarajah}, \bibinfo{person}{John Smolin}, \bibinfo{person}{Jay~M Gambetta}, {et~al\mbox{.}}} \bibinfo{year}{2022}\natexlab{}.
\newblock \showarticletitle{OpenQASM 3: A broader and deeper quantum assembly language}.
\newblock \bibinfo{journal}{\emph{ACM Transactions on Quantum Computing}} \bibinfo{volume}{3}, \bibinfo{number}{3} (\bibinfo{year}{2022}), \bibinfo{pages}{1--50}.
\newblock
\href{https://doi.org/10.1145/3505636}{doi:\nolinkurl{10.1145/3505636}}


\bibitem[Deutsch(1985)]%
        {Deutsch1985}
\bibfield{author}{\bibinfo{person}{D. Deutsch}.} \bibinfo{year}{1985}\natexlab{}.
\newblock \showarticletitle{Quantum theory, the Church–Turing principle and the universal quantum computer}.
\newblock \bibinfo{journal}{\emph{Proceedings of the Royal Society of London. A. Mathematical and Physical Sciences}}  \bibinfo{volume}{400} (\bibinfo{year}{1985}), \bibinfo{pages}{97--117}.
\newblock
Issue 1818.
\showISSN{00804630}
\href{https://doi.org/10.1098/RSPA.1985.0070}{doi:\nolinkurl{10.1098/RSPA.1985.0070}}


\bibitem[D{\'\i}az et~al\mbox{.}(2025)]%
        {Diaz2025}
\bibfield{author}{\bibinfo{person}{Ana D{\'\i}az}, \bibinfo{person}{Jaime Alvarado-Valiente}, \bibinfo{person}{Javier Romero-{\'A}lvarez}, \bibinfo{person}{Enrique Moguel}, \bibinfo{person}{Jose Garcia-Alonso}, \bibinfo{person}{Mois{\'e}s Rodr{\'\i}guez}, \bibinfo{person}{Ignacio Garc{\'\i}a-Rodr{\'\i}guez}, {and} \bibinfo{person}{Juan~M Murillo}.} \bibinfo{year}{2025}\natexlab{}.
\newblock \showarticletitle{Service engineering for quantum computing: Ensuring high-quality quantum services}.
\newblock \bibinfo{journal}{\emph{Information and Software Technology}}  \bibinfo{volume}{179} (\bibinfo{year}{2025}), \bibinfo{pages}{1--13}.
\newblock
\href{https://doi.org/10.1016/j.infsof.2024.107643}{doi:\nolinkurl{10.1016/j.infsof.2024.107643}}


\bibitem[Falkenthal et~al\mbox{.}(2024)]%
        {Falkenthal2024}
\bibfield{author}{\bibinfo{person}{Michael Falkenthal}, \bibinfo{person}{Christoph Krieger}, \bibinfo{person}{Felix Paul}, \bibinfo{person}{Sebastian Wagner}, {and} \bibinfo{person}{Michael Wurster}.} \bibinfo{year}{2024}\natexlab{}.
\newblock \showarticletitle{Planqk—platform and ecosystem for quantum applications}.
\newblock \bibinfo{journal}{\emph{KI-K{\"u}nstliche Intelligenz}} \bibinfo{volume}{38}, \bibinfo{number}{4} (\bibinfo{year}{2024}), \bibinfo{pages}{371--377}.
\newblock
\href{https://doi.org/10.1007/s13218-024-00865-6}{doi:\nolinkurl{10.1007/s13218-024-00865-6}}


\bibitem[Faro et~al\mbox{.}(2023)]%
        {Faro2023}
\bibfield{author}{\bibinfo{person}{Ismael Faro}, \bibinfo{person}{Iskandar Sitdikov}, \bibinfo{person}{David~Garcia Valiñas}, \bibinfo{person}{Francisco Jose~Martin Fernandez}, \bibinfo{person}{Christopher Codella}, {and} \bibinfo{person}{Jennifer Glick}.} \bibinfo{year}{2023}\natexlab{}.
\newblock \showarticletitle{Middleware for Quantum: An orchestration of hybrid quantum-classical systems}. In \bibinfo{booktitle}{\emph{2023 IEEE International Conference on Quantum Software (QSW)}}. \bibinfo{publisher}{IEEE}, \bibinfo{address}{Chicago, IL, USA}, \bibinfo{pages}{1--8}.
\newblock
\href{https://doi.org/10.1109/QSW59989.2023.00011}{doi:\nolinkurl{10.1109/QSW59989.2023.00011}}


\bibitem[Feynman(1982)]%
        {Feynman1982}
\bibfield{author}{\bibinfo{person}{Richard~P. Feynman}.} \bibinfo{year}{1982}\natexlab{}.
\newblock \showarticletitle{Simulating physics with computers}.
\newblock \bibinfo{journal}{\emph{International Journal of Theoretical Physics}} \bibinfo{volume}{21}, \bibinfo{number}{6} (\bibinfo{year}{1982}), \bibinfo{pages}{467--488}.
\newblock
\showISBNx{1572-9575}
\href{https://doi.org/10.1007/BF02650179}{doi:\nolinkurl{10.1007/BF02650179}}


\bibitem[Fresno-Aranda et~al\mbox{.}(2022)]%
        {fresno2022}
\bibfield{author}{\bibinfo{person}{Rafael Fresno-Aranda}, \bibinfo{person}{Pablo Fern{\'a}ndez}, \bibinfo{person}{Amador Dur{\'a}n}, {and} \bibinfo{person}{Antonio Ruiz-Cort{\'e}s}.} \bibinfo{year}{2022}\natexlab{}.
\newblock \showarticletitle{Semi-automated capacity analysis of limitation-aware microservices architectures}. In \bibinfo{booktitle}{\emph{International Conference on the Economics of Grids, Clouds, Systems, and Services}}. Springer, \bibinfo{pages}{75--88}.
\newblock
\href{https://doi.org/10.1007/978-3-031-29315-3_7}{doi:\nolinkurl{10.1007/978-3-031-29315-3_7}}


\bibitem[Gamez-Diaz et~al\mbox{.}(2018)]%
        {GamezDiaz2018}
\bibfield{author}{\bibinfo{person}{Antonio Gamez-Diaz}, \bibinfo{person}{Pablo Fernandez}, {and} \bibinfo{person}{Antonio Ruiz-Cortes}.} \bibinfo{year}{2018}\natexlab{}.
\newblock \showarticletitle{SLA-Driven Governance for RESTful Systems}. In \bibinfo{booktitle}{\emph{Service-Oriented Computing, ICSOC 2017 Workshops}}, \bibfield{editor}{\bibinfo{person}{Lars Braubach}, \bibinfo{person}{Juan~M. Murillo}, \bibinfo{person}{Nima Kaviani}, \bibinfo{person}{Manuel Lama}, \bibinfo{person}{Loli Burgue{\~{n}}o}, \bibinfo{person}{Naouel Moha}, {and} \bibinfo{person}{Marc Oriol}} (Eds.). \bibinfo{publisher}{Springer}, \bibinfo{address}{Cham}, \bibinfo{pages}{352--356}.
\newblock
\showISBNx{978-3-319-91764-1}
\href{https://doi.org/10.1007/978-3-319-91764-1_30}{doi:\nolinkurl{10.1007/978-3-319-91764-1_30}}


\bibitem[Garcia-Alonso et~al\mbox{.}(2021)]%
        {Garcia2021}
\bibfield{author}{\bibinfo{person}{Jose Garcia-Alonso}, \bibinfo{person}{Javier Rojo}, \bibinfo{person}{David Valencia}, \bibinfo{person}{Enrique Moguel}, \bibinfo{person}{Javier Berrocal}, {and} \bibinfo{person}{Juan~Manuel Murillo}.} \bibinfo{year}{2021}\natexlab{}.
\newblock \showarticletitle{Quantum software as a service through a quantum API gateway}.
\newblock \bibinfo{journal}{\emph{IEEE Internet Computing}} \bibinfo{volume}{26}, \bibinfo{number}{1} (\bibinfo{year}{2021}), \bibinfo{pages}{34--41}.
\newblock
\href{https://doi.org/10.1109/MIC.2021.3132688}{doi:\nolinkurl{10.1109/MIC.2021.3132688}}


\bibitem[Garc{\'\i}a-Fern{\'a}ndez et~al\mbox{.}(2024)]%
        {GarciaFernandez2024}
\bibfield{author}{\bibinfo{person}{Alejandro Garc{\'\i}a-Fern{\'a}ndez}, \bibinfo{person}{Jos{\'e}~Antonio Parejo}, \bibinfo{person}{Francisco~Javier Cavero}, {and} \bibinfo{person}{Antonio Ruiz-Cort{\'e}s}.} \bibinfo{year}{2024}\natexlab{}.
\newblock \showarticletitle{Racing the Market: An Industry Support Analysis for Pricing-Driven DevOps in SaaS}. In \bibinfo{booktitle}{\emph{International Conference on Service-Oriented Computing}}. \bibinfo{publisher}{Springer}, \bibinfo{address}{Cham}, \bibinfo{pages}{260--275}.
\newblock
\href{https://doi.org/10.1007/978-981-96-0808-9_19}{doi:\nolinkurl{10.1007/978-981-96-0808-9_19}}


\bibitem[Garc{\'i}a-Fern{\'a}ndez et~al\mbox{.}(2025)]%
        {GarciaFernandez2025}
\bibfield{author}{\bibinfo{person}{Alejandro Garc{\'i}a-Fern{\'a}ndez}, \bibinfo{person}{Jos{\'e}~Antonio Parejo}, \bibinfo{person}{Pablo Trinidad}, {and} \bibinfo{person}{Antonio Ruiz-Cort{\'e}s}.} \bibinfo{year}{2025}\natexlab{}.
\newblock \showarticletitle{Automated Analysis of Pricings in SaaS-Based Information Systems}. In \bibinfo{booktitle}{\emph{Advanced Information Systems Engineering}}, \bibfield{editor}{\bibinfo{person}{John Krogstie}, \bibinfo{person}{Stefanie Rinderle-Ma}, \bibinfo{person}{Gerti Kappel}, {and} \bibinfo{person}{Henderik~A. Proper}} (Eds.). \bibinfo{publisher}{Springer}, \bibinfo{address}{Cham}, \bibinfo{pages}{223--239}.
\newblock
\showISBNx{978-3-031-94571-7}


\bibitem[Grossi et~al\mbox{.}(2021)]%
        {Grossi2021}
\bibfield{author}{\bibinfo{person}{Michele Grossi}, \bibinfo{person}{Luca Crippa}, \bibinfo{person}{Antonello Aita}, \bibinfo{person}{Giacomo Bartoli}, \bibinfo{person}{Vito Sammarco}, \bibinfo{person}{Eleonora Picca}, \bibinfo{person}{Najla Said}, \bibinfo{person}{Filippo Tramonto}, {and} \bibinfo{person}{Federico Mattei}.} \bibinfo{year}{2021}\natexlab{}.
\newblock \showarticletitle{A serverless cloud integration for quantum computing}.
\newblock \bibinfo{journal}{\emph{arXiv preprint arXiv:2107.02007}} \bibinfo{volume}{1}, \bibinfo{number}{1} (\bibinfo{year}{2021}), \bibinfo{pages}{1--8}.
\newblock
\href{https://doi.org/10.48550/arXiv.2107.02007}{doi:\nolinkurl{10.48550/arXiv.2107.02007}}


\bibitem[Grover(1996)]%
        {Grover1996}
\bibfield{author}{\bibinfo{person}{Lov~K. Grover}.} \bibinfo{year}{1996}\natexlab{}.
\newblock \showarticletitle{A fast quantum mechanical algorithm for database search}. In \bibinfo{booktitle}{\emph{Proceedings of the Twenty-Eighth Annual ACM Symposium on Theory of Computing}} (Philadelphia, Pennsylvania, USA) \emph{(\bibinfo{series}{STOC '96})}. \bibinfo{publisher}{Association for Computing Machinery}, \bibinfo{address}{New York, NY, USA}, \bibinfo{pages}{212--–219}.
\newblock
\showISBNx{0897917855}
\href{https://doi.org/10.1145/237814.237866}{doi:\nolinkurl{10.1145/237814.237866}}


\bibitem[Haghparast et~al\mbox{.}(2024)]%
        {Haghparast2024}
\bibfield{author}{\bibinfo{person}{Majid Haghparast}, \bibinfo{person}{Enrique Moguel}, \bibinfo{person}{Jose Garcia-Alonso}, \bibinfo{person}{Tommi Mikkonen}, {and} \bibinfo{person}{Juan~Manuel Murillo}.} \bibinfo{year}{2024}\natexlab{}.
\newblock \showarticletitle{Innovative Approaches to Teaching Quantum Computer Programming and Quantum Software Engineering}.
\newblock \bibinfo{journal}{\emph{Proceedings - IEEE Quantum Week 2024, QCE 2024}}  \bibinfo{volume}{2} (\bibinfo{year}{2024}), \bibinfo{pages}{251--255}.
\newblock
\showISBNx{9798331541378}
\href{https://doi.org/10.1109/QCE60285.2024.10287}{doi:\nolinkurl{10.1109/QCE60285.2024.10287}}


\bibitem[Heng et~al\mbox{.}(2022)]%
        {Heng2022}
\bibfield{author}{\bibinfo{person}{Sovanmonynuth Heng}, \bibinfo{person}{Dongmin Kim}, \bibinfo{person}{Taekyung Kim}, {and} \bibinfo{person}{Youngsun Han}.} \bibinfo{year}{2022}\natexlab{}.
\newblock \showarticletitle{How to Solve Combinatorial Optimization Problems Using Real Quantum Machines: A Recent Survey}.
\newblock \bibinfo{journal}{\emph{IEEE Access}}  \bibinfo{volume}{10} (\bibinfo{year}{2022}), \bibinfo{pages}{120106--120121}.
\newblock
\href{https://doi.org/10.1109/ACCESS.2022.3218908}{doi:\nolinkurl{10.1109/ACCESS.2022.3218908}}


\bibitem[Irie et~al\mbox{.}(2021)]%
        {Irie2021}
\bibfield{author}{\bibinfo{person}{Hirotaka Irie}, \bibinfo{person}{Haozhao Liang}, \bibinfo{person}{Takumi Doi}, \bibinfo{person}{Shinya Gongyo}, {and} \bibinfo{person}{Tetsuo Hatsuda}.} \bibinfo{year}{2021}\natexlab{}.
\newblock \showarticletitle{Hybrid quantum annealing via molecular dynamics}.
\newblock \bibinfo{journal}{\emph{Scientific reports}} \bibinfo{volume}{11}, \bibinfo{number}{1} (\bibinfo{year}{2021}), \bibinfo{pages}{1--9}.
\newblock
\href{https://doi.org/10.1038/s41598-021-87676-z}{doi:\nolinkurl{10.1038/s41598-021-87676-z}}


\bibitem[Jatoth and Gangadharan(2015)]%
        {Jatoth2015}
\bibfield{author}{\bibinfo{person}{Chandrashekar Jatoth} {and} \bibinfo{person}{G.~R. Gangadharan}.} \bibinfo{year}{2015}\natexlab{}.
\newblock \showarticletitle{QoS-Aware Web Service Composition Using Quantum Inspired Particle Swarm Optimization}. In \bibinfo{booktitle}{\emph{Intelligent Decision Technologies}}, \bibfield{editor}{\bibinfo{person}{Rui Neves-Silva}, \bibinfo{person}{Lakhmi~C. Jain}, {and} \bibinfo{person}{Robert~J. Howlett}} (Eds.). \bibinfo{publisher}{Springer}, \bibinfo{address}{Cham}, \bibinfo{pages}{255--265}.
\newblock
\showISBNx{978-3-319-19857-6}
\href{https://doi.org/10.1007/978-3-319-19857-6_23}{doi:\nolinkurl{10.1007/978-3-319-19857-6_23}}


\bibitem[Klein(1961)]%
        {Klein1961}
\bibfield{author}{\bibinfo{person}{Martin~J Klein}.} \bibinfo{year}{1961}\natexlab{}.
\newblock \showarticletitle{Max Planck and the beginnings of the quantum theory}.
\newblock \bibinfo{journal}{\emph{Archive for History of Exact Sciences}} \bibinfo{volume}{1}, \bibinfo{number}{5} (\bibinfo{year}{1961}), \bibinfo{pages}{459--479}.
\newblock
\href{https://doi.org/10.1007/BF00327765}{doi:\nolinkurl{10.1007/BF00327765}}


\bibitem[Knill(2010)]%
        {Knill2010}
\bibfield{author}{\bibinfo{person}{Emanuel Knill}.} \bibinfo{year}{2010}\natexlab{}.
\newblock \showarticletitle{Quantum computing}.
\newblock \bibinfo{journal}{\emph{Nature}}  \bibinfo{volume}{463} (\bibinfo{year}{2010}), \bibinfo{pages}{441--443}.
\newblock
\href{https://doi.org/10.1038/463441a}{doi:\nolinkurl{10.1038/463441a}}


\bibitem[Kumara et~al\mbox{.}(2021)]%
        {Kumara2021}
\bibfield{author}{\bibinfo{person}{Indika Kumara}, \bibinfo{person}{Willem-Jan Van Den~Heuvel}, {and} \bibinfo{person}{Damian~A Tamburri}.} \bibinfo{year}{2021}\natexlab{}.
\newblock \showarticletitle{QSOC: Quantum service-oriented computing}. In \bibinfo{booktitle}{\emph{Symposium and Summer School on Service-Oriented Computing}}. \bibinfo{publisher}{Springer}, \bibinfo{address}{Cham}, \bibinfo{pages}{52--63}.
\newblock
\href{https://doi.org/10.1007/978-3-030-87568-8_3}{doi:\nolinkurl{10.1007/978-3-030-87568-8_3}}


\bibitem[Ladd et~al\mbox{.}(2010)]%
        {Ladd2010}
\bibfield{author}{\bibinfo{person}{Thaddeus~D Ladd}, \bibinfo{person}{Fedor Jelezko}, \bibinfo{person}{Raymond Laflamme}, \bibinfo{person}{Yasunobu Nakamura}, \bibinfo{person}{Christopher Monroe}, {and} \bibinfo{person}{Jeremy~Lloyd O’Brien}.} \bibinfo{year}{2010}\natexlab{}.
\newblock \showarticletitle{Quantum computers}.
\newblock \bibinfo{journal}{\emph{nature}} \bibinfo{volume}{464}, \bibinfo{number}{7285} (\bibinfo{year}{2010}), \bibinfo{pages}{45--53}.
\newblock
\href{https://doi.org/10.1038/nature08812}{doi:\nolinkurl{10.1038/nature08812}}


\bibitem[Leite~Ramalho et~al\mbox{.}(2025)]%
        {Leite2025}
\bibfield{author}{\bibinfo{person}{Neilson~Carlos Leite~Ramalho}, \bibinfo{person}{Higor Amario~de Souza}, {and} \bibinfo{person}{Marcos Lordello~Chaim}.} \bibinfo{year}{2025}\natexlab{}.
\newblock \showarticletitle{Testing and Debugging Quantum Programs: The Road to 2030}.
\newblock \bibinfo{journal}{\emph{ACM Trans. Softw. Eng. Methodol.}} \bibinfo{volume}{34}, \bibinfo{number}{5} (\bibinfo{year}{2025}), \bibinfo{pages}{1--46}.
\newblock
\showISSN{1049-331X}
\href{https://doi.org/10.1145/3715106}{doi:\nolinkurl{10.1145/3715106}}


\bibitem[Leymann(2019)]%
        {Leymann2019}
\bibfield{author}{\bibinfo{person}{Frank Leymann}.} \bibinfo{year}{2019}\natexlab{}.
\newblock \showarticletitle{Towards a pattern language for quantum algorithms}. In \bibinfo{booktitle}{\emph{Quantum Technology and Optimization Problems: First International Workshop, QTOP 2019, Munich, Germany, March 18, 2019, Proceedings 1}}. \bibinfo{publisher}{Springer}, \bibinfo{address}{Cham}, \bibinfo{pages}{218--230}.
\newblock
\href{https://doi.org/10.1007/978-3-030-14082-3_19}{doi:\nolinkurl{10.1007/978-3-030-14082-3_19}}


\bibitem[Leymann and Barzen(2020)]%
        {Leymann2020}
\bibfield{author}{\bibinfo{person}{Frank Leymann} {and} \bibinfo{person}{Johanna Barzen}.} \bibinfo{year}{2020}\natexlab{}.
\newblock \showarticletitle{The bitter truth about gate-based quantum algorithms in the NISQ era}.
\newblock \bibinfo{journal}{\emph{Quantum Science and Technology}} \bibinfo{volume}{5}, \bibinfo{number}{4} (\bibinfo{year}{2020}), \bibinfo{pages}{1--29}.
\newblock
\href{https://doi.org/10.1088/2058-9565/abae7d}{doi:\nolinkurl{10.1088/2058-9565/abae7d}}


\bibitem[Luo and Zhao(2025)]%
        {Luo2025}
\bibfield{author}{\bibinfo{person}{Junjie Luo} {and} \bibinfo{person}{Jianjun Zhao}.} \bibinfo{year}{2025}\natexlab{}.
\newblock \showarticletitle{Formalization of quantum intermediate representations for code safety}.
\newblock \bibinfo{journal}{\emph{Journal of Systems and Software}}  \bibinfo{volume}{219} (\bibinfo{year}{2025}), \bibinfo{pages}{1--12}.
\newblock
\href{https://doi.org/10.1016/j.jss.2024.112236}{doi:\nolinkurl{10.1016/j.jss.2024.112236}}


\bibitem[McCaskey et~al\mbox{.}(2020)]%
        {Mccaskey2020}
\bibfield{author}{\bibinfo{person}{Alexander~J McCaskey}, \bibinfo{person}{Dmitry~I Lyakh}, \bibinfo{person}{Eugene~F Dumitrescu}, \bibinfo{person}{Sarah~S Powers}, {and} \bibinfo{person}{Travis~S Humble}.} \bibinfo{year}{2020}\natexlab{}.
\newblock \showarticletitle{XACC: a system-level software infrastructure for heterogeneous quantum--classical computing}.
\newblock \bibinfo{journal}{\emph{Quantum Science and Technology}} \bibinfo{volume}{5}, \bibinfo{number}{2} (\bibinfo{year}{2020}), \bibinfo{pages}{1--24}.
\newblock
\href{https://doi.org/10.1088/2058-9565/ab6bf6}{doi:\nolinkurl{10.1088/2058-9565/ab6bf6}}


\bibitem[McClean et~al\mbox{.}(2016)]%
        {McClean2016}
\bibfield{author}{\bibinfo{person}{Jarrod~R McClean}, \bibinfo{person}{Jonathan Romero}, \bibinfo{person}{Ryan Babbush}, {and} \bibinfo{person}{Alán Aspuru-Guzik}.} \bibinfo{year}{2016}\natexlab{}.
\newblock \showarticletitle{The theory of variational hybrid quantum-classical algorithms}.
\newblock \bibinfo{journal}{\emph{New Journal of Physics}} \bibinfo{volume}{18}, \bibinfo{number}{2} (\bibinfo{year}{2016}), \bibinfo{pages}{1--23}.
\newblock
\href{https://doi.org/10.1088/1367-2630/18/2/023023}{doi:\nolinkurl{10.1088/1367-2630/18/2/023023}}


\bibitem[Moguel et~al\mbox{.}(2022)]%
        {Moguel2022}
\bibfield{author}{\bibinfo{person}{Enrique Moguel}, \bibinfo{person}{Javier Rojo}, \bibinfo{person}{David Valencia}, \bibinfo{person}{Javier Berrocal}, \bibinfo{person}{Jose Garcia-Alonso}, {and} \bibinfo{person}{Juan~M Murillo}.} \bibinfo{year}{2022}\natexlab{}.
\newblock \showarticletitle{Quantum service-oriented computing: current landscape and challenges}.
\newblock \bibinfo{journal}{\emph{Software Quality Journal}} \bibinfo{volume}{30}, \bibinfo{number}{4} (\bibinfo{year}{2022}), \bibinfo{pages}{983--1002}.
\newblock
\href{https://doi.org/10.1007/S11219-022-09589-Y}{doi:\nolinkurl{10.1007/S11219-022-09589-Y}}


\bibitem[Muqeet et~al\mbox{.}(2024)]%
        {Muqeet2024}
\bibfield{author}{\bibinfo{person}{Asmar Muqeet}, \bibinfo{person}{Tao Yue}, \bibinfo{person}{Shaukat Ali}, {and} \bibinfo{person}{Paolo Arcaini}.} \bibinfo{year}{2024}\natexlab{}.
\newblock \showarticletitle{Mitigating Noise in Quantum Software Testing Using Machine Learning}.
\newblock \bibinfo{journal}{\emph{IEEE Transactions on Software Engineering}} \bibinfo{volume}{50}, \bibinfo{number}{11} (\bibinfo{year}{2024}), \bibinfo{pages}{2947--2961}.
\newblock
\href{https://doi.org/10.1109/TSE.2024.3462974}{doi:\nolinkurl{10.1109/TSE.2024.3462974}}


\bibitem[Murillo et~al\mbox{.}(2025)]%
        {Murillo2025}
\bibfield{author}{\bibinfo{person}{Juan~M Murillo}, \bibinfo{person}{Jose Garcia-Alonso}, \bibinfo{person}{Enrique Moguel}, \bibinfo{person}{Johanna Barzen}, \bibinfo{person}{Frank Leymann}, \bibinfo{person}{Shaukat Ali}, \bibinfo{person}{Tao Yue}, \bibinfo{person}{Paolo Arcaini}, \bibinfo{person}{Ricardo P{\'e}rez-Castillo}, \bibinfo{person}{Ignacio Garc{\'\i}a Rodr{\'\i}guez~de Guzm{\'a}n}, \bibinfo{person}{Mario Piattini}, \bibinfo{person}{Antonio Ruiz-Cortes}, \bibinfo{person}{Antonio Brogi}, \bibinfo{person}{Jianjun Zhao}, \bibinfo{person}{Andriy Miranskyy}, {and} \bibinfo{person}{Manuel Wimmer}.} \bibinfo{year}{2025}\natexlab{}.
\newblock \showarticletitle{Quantum Software Engineering: Roadmap and Challenges Ahead}.
\newblock \bibinfo{journal}{\emph{ACM Transactions on Software Engineering and Methodology}} \bibinfo{volume}{34}, \bibinfo{number}{5} (\bibinfo{year}{2025}), \bibinfo{pages}{1--48}.
\newblock
\href{https://doi.org/10.1145/3712002}{doi:\nolinkurl{10.1145/3712002}}


\bibitem[Nguyen et~al\mbox{.}(2024a)]%
        {Nguyen2024cloud}
\bibfield{author}{\bibinfo{person}{Hoa~T Nguyen}, \bibinfo{person}{Prabhakar Krishnan}, \bibinfo{person}{Dilip Krishnaswamy}, \bibinfo{person}{Muhammad Usman}, {and} \bibinfo{person}{Rajkumar Buyya}.} \bibinfo{year}{2024}\natexlab{a}.
\newblock \showarticletitle{Quantum Cloud Computing: A Review, Open Problems, and Future Directions}.
\newblock \bibinfo{journal}{\emph{arXiv preprint arXiv:2404.11420}} (\bibinfo{year}{2024}).
\newblock
\href{https://doi.org/10.48550/arXiv.2404.11420}{doi:\nolinkurl{10.48550/arXiv.2404.11420}}


\bibitem[Nguyen et~al\mbox{.}(2024b)]%
        {Nguyen2024}
\bibfield{author}{\bibinfo{person}{Hoa~T Nguyen}, \bibinfo{person}{Muhammad Usman}, {and} \bibinfo{person}{Rajkumar Buyya}.} \bibinfo{year}{2024}\natexlab{b}.
\newblock \showarticletitle{Qfaas: A serverless function-as-a-service framework for quantum computing}.
\newblock \bibinfo{journal}{\emph{Future Generation Computer Systems}}  \bibinfo{volume}{154} (\bibinfo{year}{2024}), \bibinfo{pages}{281--300}.
\newblock
\href{https://doi.org/10.1016/j.future.2024.01.018}{doi:\nolinkurl{10.1016/j.future.2024.01.018}}


\bibitem[Nielsen and Chuang(2010)]%
        {Nielsen2010}
\bibfield{author}{\bibinfo{person}{Michael~A Nielsen} {and} \bibinfo{person}{Isaac~L Chuang}.} \bibinfo{year}{2010}\natexlab{}.
\newblock \bibinfo{booktitle}{\emph{Quantum computation and quantum information}}.
\newblock \bibinfo{publisher}{Cambridge university press}, \bibinfo{address}{Cambridge}.
\newblock
\href{https://doi.org/10.1017/CBO9780511976667}{doi:\nolinkurl{10.1017/CBO9780511976667}}


\bibitem[Orus et~al\mbox{.}(2019)]%
        {Orus2019}
\bibfield{author}{\bibinfo{person}{Roman Orus}, \bibinfo{person}{Samuel Mugel}, {and} \bibinfo{person}{Enrique Lizaso}.} \bibinfo{year}{2019}\natexlab{}.
\newblock \showarticletitle{Quantum computing for finance: overview and prospects}.
\newblock \bibinfo{journal}{\emph{Reviews in Physics}}  \bibinfo{volume}{4} (\bibinfo{year}{2019}), \bibinfo{pages}{1--12}.
\newblock
\href{https://doi.org/10.1016/j.revip.2019.100028}{doi:\nolinkurl{10.1016/j.revip.2019.100028}}


\bibitem[O’Riordan et~al\mbox{.}(2020)]%
        {ORiordan2020}
\bibfield{author}{\bibinfo{person}{Lee~J O’Riordan}, \bibinfo{person}{Myles Doyle}, \bibinfo{person}{Fabio Baruffa}, {and} \bibinfo{person}{Venkatesh Kannan}.} \bibinfo{year}{2020}\natexlab{}.
\newblock \showarticletitle{A hybrid classical-quantum workflow for natural language processing}.
\newblock \bibinfo{journal}{\emph{Machine Learning: Science and Technology}} \bibinfo{volume}{2}, \bibinfo{number}{1} (\bibinfo{year}{2020}), \bibinfo{pages}{1--25}.
\newblock
\href{https://doi.org/10.1088/2632-2153/abbd2e}{doi:\nolinkurl{10.1088/2632-2153/abbd2e}}


\bibitem[Peelam et~al\mbox{.}(2024)]%
        {Peelam2024}
\bibfield{author}{\bibinfo{person}{Mritunjay~Shall Peelam}, \bibinfo{person}{Anjaney~Asreet Rout}, {and} \bibinfo{person}{Vinay Chamola}.} \bibinfo{year}{2024}\natexlab{}.
\newblock \showarticletitle{Quantum computing applications for Internet of Things}.
\newblock \bibinfo{journal}{\emph{IET Quantum Communication}} \bibinfo{volume}{5}, \bibinfo{number}{2} (\bibinfo{year}{2024}), \bibinfo{pages}{103--112}.
\newblock
\href{https://doi.org/10.1049/qtc2.12079}{doi:\nolinkurl{10.1049/qtc2.12079}}


\bibitem[P{\'e}rez-Castillo and Piattini(2022)]%
        {Perez2022}
\bibfield{author}{\bibinfo{person}{Ricardo P{\'e}rez-Castillo} {and} \bibinfo{person}{Mario Piattini}.} \bibinfo{year}{2022}\natexlab{}.
\newblock \showarticletitle{Design of classical-quantum systems with UML}.
\newblock \bibinfo{journal}{\emph{Computing}} \bibinfo{volume}{104}, \bibinfo{number}{11} (\bibinfo{year}{2022}), \bibinfo{pages}{2375--2403}.
\newblock
\href{https://doi.org/10.1007/s00607-022-01091-4}{doi:\nolinkurl{10.1007/s00607-022-01091-4}}


\bibitem[Petersen(1963)]%
        {Petersen1963}
\bibfield{author}{\bibinfo{person}{Aage Petersen}.} \bibinfo{year}{1963}\natexlab{}.
\newblock \showarticletitle{The philosophy of niels bohr}.
\newblock \bibinfo{journal}{\emph{Bulletin of the atomic scientists}} \bibinfo{volume}{19}, \bibinfo{number}{7} (\bibinfo{year}{1963}), \bibinfo{pages}{8--14}.
\newblock
\href{https://doi.org/10.1080/00963402.1963.11454520}{doi:\nolinkurl{10.1080/00963402.1963.11454520}}


\bibitem[Piattini et~al\mbox{.}(2020)]%
        {Piattini2020}
\bibfield{author}{\bibinfo{person}{Mario Piattini}, \bibinfo{person}{Guido Peterssen}, \bibinfo{person}{Ricardo P{\'e}rez-Castillo}, \bibinfo{person}{Jose~Luis Hevia}, \bibinfo{person}{Manuel~A Serrano}, \bibinfo{person}{Guillermo Hern{\'a}ndez}, \bibinfo{person}{Ignacio Garc{\'\i}a~Rodr{\'\i}guez De~Guzm{\'a}n}, \bibinfo{person}{Claudio~Andr{\'e}s Paradela}, \bibinfo{person}{Macario Polo}, \bibinfo{person}{Ezequiel Murina}, {et~al\mbox{.}}} \bibinfo{year}{2020}\natexlab{}.
\newblock \showarticletitle{The Talavera Manifesto for quantum software engineering and programming.}. In \bibinfo{booktitle}{\emph{QANSWER}}. \bibinfo{publisher}{CEUR}, \bibinfo{address}{CEUR-WS.org}, \bibinfo{pages}{1--5}.
\newblock
\urldef\tempurl%
\url{https://ceur-ws.org/Vol-2561/paper0.pdf}
\showURL{%
\tempurl}


\bibitem[Preskill(2018)]%
        {Preskill2018}
\bibfield{author}{\bibinfo{person}{John Preskill}.} \bibinfo{year}{2018}\natexlab{}.
\newblock \showarticletitle{Quantum {C}omputing in the {NISQ} era and beyond}.
\newblock \bibinfo{journal}{\emph{{Quantum}}}  \bibinfo{volume}{2} (\bibinfo{year}{2018}), \bibinfo{pages}{1--20}.
\newblock
\showISSN{2521-327X}
\href{https://doi.org/10.22331/q-2018-08-06-79}{doi:\nolinkurl{10.22331/q-2018-08-06-79}}


\bibitem[Quetschlich et~al\mbox{.}(2023)]%
        {Quetschlich2023}
\bibfield{author}{\bibinfo{person}{Nils Quetschlich}, \bibinfo{person}{Lukas Burgholzer}, {and} \bibinfo{person}{Robert Wille}.} \bibinfo{year}{2023}\natexlab{}.
\newblock \showarticletitle{Predicting Good Quantum Circuit Compilation Options}. In \bibinfo{booktitle}{\emph{2023 IEEE International Conference on Quantum Software (QSW)}}. \bibinfo{publisher}{IEEE}, \bibinfo{address}{Chicago, IL, USA}, \bibinfo{pages}{43--53}.
\newblock
\href{https://doi.org/10.1109/QSW59989.2023.00015}{doi:\nolinkurl{10.1109/QSW59989.2023.00015}}


\bibitem[Qui\~{n}a Mera et~al\mbox{.}(2023)]%
        {Quinamera2023}
\bibfield{author}{\bibinfo{person}{Antonio Qui\~{n}a Mera}, \bibinfo{person}{Pablo Fernandez}, \bibinfo{person}{Jos\'{e}~Mar\'{\i}a Garc\'{\i}a}, {and} \bibinfo{person}{Antonio Ruiz-Cort\'{e}s}.} \bibinfo{year}{2023}\natexlab{}.
\newblock \showarticletitle{GraphQL: A Systematic Mapping Study}.
\newblock \bibinfo{journal}{\emph{ACM Comput. Surv.}} \bibinfo{volume}{55}, \bibinfo{number}{10} (\bibinfo{year}{2023}), \bibinfo{pages}{1--35}.
\newblock
\showISSN{0360-0300}
\href{https://doi.org/10.1145/3561818}{doi:\nolinkurl{10.1145/3561818}}


\bibitem[Ravi et~al\mbox{.}(2021)]%
        {Ravi2021}
\bibfield{author}{\bibinfo{person}{Gokul~Subramanian Ravi}, \bibinfo{person}{Kaitlin~N. Smith}, \bibinfo{person}{Prakash Murali}, {and} \bibinfo{person}{Frederic~T. Chong}.} \bibinfo{year}{2021}\natexlab{}.
\newblock \showarticletitle{Adaptive job and resource management for the growing quantum cloud}. In \bibinfo{booktitle}{\emph{2021 IEEE International Conference on Quantum Computing and Engineering (QCE)}}. \bibinfo{publisher}{IEEE}, \bibinfo{address}{Broomfield, CO, USA}, \bibinfo{pages}{301--312}.
\newblock
\href{https://doi.org/10.1109/QCE52317.2021.00047}{doi:\nolinkurl{10.1109/QCE52317.2021.00047}}


\bibitem[Riedel et~al\mbox{.}(2019)]%
        {Riedel2019}
\bibfield{author}{\bibinfo{person}{Max Riedel}, \bibinfo{person}{Matyas Kovacs}, \bibinfo{person}{Peter Zoller}, \bibinfo{person}{J{\"u}rgen Mlynek}, {and} \bibinfo{person}{Tommaso Calarco}.} \bibinfo{year}{2019}\natexlab{}.
\newblock \showarticletitle{Europe’s quantum flagship initiative}.
\newblock \bibinfo{journal}{\emph{Quantum Science and Technology}} \bibinfo{volume}{4}, \bibinfo{number}{2} (\bibinfo{year}{2019}), \bibinfo{pages}{1--7}.
\newblock
\href{https://doi.org/10.1088/2058-9565/ab042d}{doi:\nolinkurl{10.1088/2058-9565/ab042d}}


\bibitem[Romero-{\'A}lvarez et~al\mbox{.}(2024a)]%
        {Romero2024c}
\bibfield{author}{\bibinfo{person}{Javier Romero-{\'A}lvarez}, \bibinfo{person}{Jaime Alvarado-Valiente}, \bibinfo{person}{Jorge Casco-Seco}, \bibinfo{person}{Enrique Moguel}, \bibinfo{person}{Jose Garcia-Alonso}, {and} \bibinfo{person}{Juan~M Murillo}.} \bibinfo{year}{2024}\natexlab{a}.
\newblock \showarticletitle{A noise validation for quantum circuit scheduling through a service-oriented architecture}.
\newblock \bibinfo{journal}{\emph{International Journal of Software Engineering and Knowledge Engineering}} \bibinfo{volume}{34}, \bibinfo{number}{09} (\bibinfo{year}{2024}), \bibinfo{pages}{1371--1386}.
\newblock
\href{https://doi.org/10.1142/s0218194024410018}{doi:\nolinkurl{10.1142/s0218194024410018}}


\bibitem[Romero-{\'A}lvarez et~al\mbox{.}(2021)]%
        {Romero2021}
\bibfield{author}{\bibinfo{person}{Javier Romero-{\'A}lvarez}, \bibinfo{person}{Jaime Alvarado-Valiente}, \bibinfo{person}{Jose Garcia-Alonso}, \bibinfo{person}{Enrique Moguel}, {and} \bibinfo{person}{Juan~M Murillo}.} \bibinfo{year}{2021}\natexlab{}.
\newblock \showarticletitle{A graph-based healthcare system for quantum simulation of medication administration in the aging people}. In \bibinfo{booktitle}{\emph{International Workshop on Gerontechnology}}. \bibinfo{publisher}{Springer}, \bibinfo{address}{Cham}, \bibinfo{pages}{34--41}.
\newblock
\href{https://doi.org/10.1007/978-3-030-97524-1_4}{doi:\nolinkurl{10.1007/978-3-030-97524-1_4}}


\bibitem[Romero-{\'A}lvarez et~al\mbox{.}(2022)]%
        {romero2022using}
\bibfield{author}{\bibinfo{person}{Javier Romero-{\'A}lvarez}, \bibinfo{person}{Jaime Alvarado-Valiente}, \bibinfo{person}{Enrique Moguel}, \bibinfo{person}{Jos{\'e} Garc{\'\i}a-Alonso}, {and} \bibinfo{person}{Juan~M Murillo}.} \bibinfo{year}{2022}\natexlab{}.
\newblock \showarticletitle{Using open API for the development of hybrid classical-quantum services}. In \bibinfo{booktitle}{\emph{International conference on service-oriented computing}}. Springer, \bibinfo{pages}{364--368}.
\newblock
\href{https://doi.org/10.1007/978-3-031-26507-5_34}{doi:\nolinkurl{10.1007/978-3-031-26507-5_34}}


\bibitem[Romero-{\'A}lvarez et~al\mbox{.}(2024b)]%
        {Romero2024}
\bibfield{author}{\bibinfo{person}{Javier Romero-{\'A}lvarez}, \bibinfo{person}{Jaime Alvarado-Valiente}, \bibinfo{person}{Enrique Moguel}, \bibinfo{person}{Jose Garcia-Alonso}, {and} \bibinfo{person}{Juan~M Murillo}.} \bibinfo{year}{2024}\natexlab{b}.
\newblock \showarticletitle{Enabling continuous deployment techniques for quantum services}.
\newblock \bibinfo{journal}{\emph{Software: Practice and Experience}} \bibinfo{volume}{54}, \bibinfo{number}{8} (\bibinfo{year}{2024}), \bibinfo{pages}{1491--1515}.
\newblock
\href{https://doi.org/10.1002/spe.3326}{doi:\nolinkurl{10.1002/spe.3326}}


\bibitem[Romero-{\'A}lvarez et~al\mbox{.}(2024c)]%
        {Romero2024b}
\bibfield{author}{\bibinfo{person}{Javier Romero-{\'A}lvarez}, \bibinfo{person}{Jaime Alvarado-Valiente}, \bibinfo{person}{Enrique Moguel}, \bibinfo{person}{Jos{\'e} Garcia-Alonso}, {and} \bibinfo{person}{Juan~M Murillo}.} \bibinfo{year}{2024}\natexlab{c}.
\newblock \bibinfo{booktitle}{\emph{Quantum Service-oriented Computing: A Proposal for Quantum Software as a Service}}.
\newblock \bibinfo{publisher}{River Publishers}, \bibinfo{address}{London}.
\newblock
\href{https://doi.org/10.1201/9788770046336}{doi:\nolinkurl{10.1201/9788770046336}}


\bibitem[Romero-Álvarez et~al\mbox{.}(2023)]%
        {Romero2023}
\bibfield{author}{\bibinfo{person}{Javier Romero-Álvarez}, \bibinfo{person}{Jaime Alvarado-Valiente}, \bibinfo{person}{Enrique Moguel}, \bibinfo{person}{Carlos Canal}, \bibinfo{person}{Jose García-Alonso}, {and} \bibinfo{person}{Juan~M. Murillo}.} \bibinfo{year}{2023}\natexlab{}.
\newblock \showarticletitle{Leveraging API Specifications for Scaffolding Quantum Applications}. In \bibinfo{booktitle}{\emph{2023 IEEE International Conference on Quantum Computing and Engineering (QCE)}}, Vol.~\bibinfo{volume}{02}. \bibinfo{publisher}{IEEE}, \bibinfo{address}{Bellevue, WA, USA}, \bibinfo{pages}{187--190}.
\newblock
\href{https://doi.org/10.1109/QCE57702.2023.10208}{doi:\nolinkurl{10.1109/QCE57702.2023.10208}}


\bibitem[Ruiz-Cort\'{e}s and Parejo(2025)]%
        {RuizCortes2025}
\bibfield{author}{\bibinfo{person}{Antonio Ruiz-Cort\'{e}s} {and} \bibinfo{person}{Jos\'{e}~Antonio Parejo}.} \bibinfo{year}{2025}\natexlab{}.
\newblock \showarticletitle{An Initial Exploration of Pricing-driven Governance for Hybrid Quantum-Classical SaaS}. In \bibinfo{booktitle}{\emph{{XXIX Jornadas de Ingeniería del Software y Bases de Datos (JISBD) - Track on Quantum Computing and Quantum Software Engineering (QuantumX)}}}. \bibinfo{publisher}{SISTEDES}, \bibinfo{address}{C\'{o}rdoba, Spain}, \bibinfo{pages}{1--13}.
\newblock
\urldef\tempurl%
\url{https://hdl.handle.net/11705/JISBD/2025/91}
\showURL{%
\tempurl}


\bibitem[Salm et~al\mbox{.}(2020)]%
        {Salm2020}
\bibfield{author}{\bibinfo{person}{Marie Salm}, \bibinfo{person}{Johanna Barzen}, \bibinfo{person}{Uwe Breitenb{\"u}cher}, \bibinfo{person}{Frank Leymann}, \bibinfo{person}{Benjamin Weder}, {and} \bibinfo{person}{Karoline Wild}.} \bibinfo{year}{2020}\natexlab{}.
\newblock \showarticletitle{The NISQ analyzer: automating the selection of quantum computers for quantum algorithms}. In \bibinfo{booktitle}{\emph{Symposium and summer school on Service-Oriented Computing}}. \bibinfo{publisher}{Springer}, \bibinfo{address}{Cham}, \bibinfo{pages}{66--85}.
\newblock
\href{https://doi.org/10.1007/978-3-030-64846-6_5}{doi:\nolinkurl{10.1007/978-3-030-64846-6_5}}


\bibitem[Salm et~al\mbox{.}(2022)]%
        {Salm2022}
\bibfield{author}{\bibinfo{person}{Marie Salm}, \bibinfo{person}{Johanna Barzen}, \bibinfo{person}{Frank Leymann}, {and} \bibinfo{person}{Philipp Wundrack}.} \bibinfo{year}{2022}\natexlab{}.
\newblock \showarticletitle{Optimizing the prioritization of compiled quantum circuits by machine learning approaches}. In \bibinfo{booktitle}{\emph{Symposium and Summer School on Service-Oriented Computing}}. \bibinfo{publisher}{Springer}, \bibinfo{address}{Cham}, \bibinfo{pages}{161--181}.
\newblock
\href{https://doi.org/10.1007/978-3-031-18304-1_9}{doi:\nolinkurl{10.1007/978-3-031-18304-1_9}}


\bibitem[Serrano et~al\mbox{.}(2022)]%
        {Serrano2022}
\bibfield{author}{\bibinfo{person}{Manuel~A Serrano}, \bibinfo{person}{Jos{\'e}~A Cruz-Lemus}, \bibinfo{person}{Ricardo Perez-Castillo}, {and} \bibinfo{person}{Mario Piattini}.} \bibinfo{year}{2022}\natexlab{}.
\newblock \showarticletitle{Quantum software components and platforms: Overview and quality assessment}.
\newblock \bibinfo{journal}{\emph{Comput. Surveys}} \bibinfo{volume}{55}, \bibinfo{number}{8} (\bibinfo{year}{2022}), \bibinfo{pages}{1--31}.
\newblock
\href{https://doi.org/10.1145/3548679}{doi:\nolinkurl{10.1145/3548679}}


\bibitem[Shor(1997)]%
        {Shor1997}
\bibfield{author}{\bibinfo{person}{Peter~W. Shor}.} \bibinfo{year}{1997}\natexlab{}.
\newblock \showarticletitle{Polynomial-Time Algorithms for Prime Factorization and Discrete Logarithms on a Quantum Computer}.
\newblock \bibinfo{journal}{\emph{SIAM J. Comput.}} \bibinfo{volume}{26}, \bibinfo{number}{5} (\bibinfo{year}{1997}), \bibinfo{pages}{1484--1509}.
\newblock
\href{https://doi.org/10.1137/S0097539795293172}{doi:\nolinkurl{10.1137/S0097539795293172}}


\bibitem[Sivarajah et~al\mbox{.}(2020)]%
        {Sivarajah2020}
\bibfield{author}{\bibinfo{person}{Seyon Sivarajah}, \bibinfo{person}{Silas Dilkes}, \bibinfo{person}{Alexander Cowtan}, \bibinfo{person}{Will Simmons}, \bibinfo{person}{Alec Edgington}, {and} \bibinfo{person}{Ross Duncan}.} \bibinfo{year}{2020}\natexlab{}.
\newblock \showarticletitle{t| ket>: a retargetable compiler for NISQ devices}.
\newblock \bibinfo{journal}{\emph{Quantum Science and Technology}} \bibinfo{volume}{6}, \bibinfo{number}{1} (\bibinfo{year}{2020}), \bibinfo{pages}{1--28}.
\newblock
\href{https://doi.org/10.1088/2058-9565/ab8e92}{doi:\nolinkurl{10.1088/2058-9565/ab8e92}}


\bibitem[Sood and Agrewal(2024)]%
        {Sood2024}
\bibfield{author}{\bibinfo{person}{Sandeep~Kumar Sood} {and} \bibinfo{person}{Monika Agrewal}.} \bibinfo{year}{2024}\natexlab{}.
\newblock \showarticletitle{Quantum machine learning for computational methods in engineering: a systematic review}.
\newblock \bibinfo{journal}{\emph{Archives of Computational Methods in Engineering}} \bibinfo{volume}{31}, \bibinfo{number}{3} (\bibinfo{year}{2024}), \bibinfo{pages}{1555--1577}.
\newblock
\href{https://doi.org/10.1007/s11831-023-10027-w}{doi:\nolinkurl{10.1007/s11831-023-10027-w}}


\bibitem[Stirbu et~al\mbox{.}(2024)]%
        {Stirbu2024}
\bibfield{author}{\bibinfo{person}{Vlad Stirbu}, \bibinfo{person}{Otso Kinanen}, \bibinfo{person}{Majid Haghparast}, {and} \bibinfo{person}{Tommi Mikkonen}.} \bibinfo{year}{2024}\natexlab{}.
\newblock \showarticletitle{Qubernetes: Towards a unified cloud-native execution platform for hybrid classic-quantum computing}.
\newblock \bibinfo{journal}{\emph{Information and Software Technology}}  \bibinfo{volume}{175} (\bibinfo{year}{2024}), \bibinfo{pages}{1--11}.
\newblock
\href{https://doi.org/10.1016/j.infsof.2024.107529}{doi:\nolinkurl{10.1016/j.infsof.2024.107529}}


\bibitem[Valencia et~al\mbox{.}(2021)]%
        {valencia2021hybrid}
\bibfield{author}{\bibinfo{person}{David Valencia}, \bibinfo{person}{Jose Garcia-Alonso}, \bibinfo{person}{Javier Rojo}, \bibinfo{person}{Enrique Moguel}, \bibinfo{person}{Javier Berrocal}, {and} \bibinfo{person}{Juan~Manuel Murillo}.} \bibinfo{year}{2021}\natexlab{}.
\newblock \showarticletitle{Hybrid classical-quantum software services systems: Exploration of the rough edges}. In \bibinfo{booktitle}{\emph{International Conference on the Quality of Information and Communications Technology}}. Springer, \bibinfo{pages}{225--238}.
\newblock
\href{https://doi.org/10.1007/978-3-030-85347-1_17}{doi:\nolinkurl{10.1007/978-3-030-85347-1_17}}


\bibitem[Weder et~al\mbox{.}(2022)]%
        {Weder2022}
\bibfield{author}{\bibinfo{person}{Benjamin Weder}, \bibinfo{person}{Johanna Barzen}, \bibinfo{person}{Frank Leymann}, {and} \bibinfo{person}{Daniel Vietz}.} \bibinfo{year}{2022}\natexlab{}.
\newblock \showarticletitle{Quantum software development lifecycle}.
\newblock In \bibinfo{booktitle}{\emph{Quantum Software Engineering}}. \bibinfo{publisher}{Springer}, \bibinfo{pages}{61--83}.
\newblock
\href{https://doi.org/10.1007/978-3-031-05324-5_4}{doi:\nolinkurl{10.1007/978-3-031-05324-5_4}}


\bibitem[Weder et~al\mbox{.}(2020)]%
        {Weder2020}
\bibfield{author}{\bibinfo{person}{Benjamin Weder}, \bibinfo{person}{Uwe Breitenb{\"u}cher}, \bibinfo{person}{Frank Leymann}, {and} \bibinfo{person}{Karoline Wild}.} \bibinfo{year}{2020}\natexlab{}.
\newblock \showarticletitle{Integrating quantum computing into workflow modeling and execution}. In \bibinfo{booktitle}{\emph{2020 IEEE/ACM 13th International Conference on Utility and Cloud Computing (UCC)}}. \bibinfo{publisher}{IEEE}, \bibinfo{address}{Leicester, UK}, \bibinfo{pages}{279--291}.
\newblock
\href{https://doi.org/10.1109/UCC48980.2020.00046}{doi:\nolinkurl{10.1109/UCC48980.2020.00046}}


\bibitem[Wei and Blake(2010)]%
        {Wei2010}
\bibfield{author}{\bibinfo{person}{Yi Wei} {and} \bibinfo{person}{M.~Brian Blake}.} \bibinfo{year}{2010}\natexlab{}.
\newblock \showarticletitle{Service-Oriented Computing and Cloud Computing: Challenges and Opportunities}.
\newblock \bibinfo{journal}{\emph{IEEE Internet Computing}} \bibinfo{volume}{14}, \bibinfo{number}{6} (\bibinfo{year}{2010}), \bibinfo{pages}{72--75}.
\newblock
\href{https://doi.org/10.1109/MIC.2010.147}{doi:\nolinkurl{10.1109/MIC.2010.147}}


\bibitem[Wild et~al\mbox{.}(2020)]%
        {Wild2020}
\bibfield{author}{\bibinfo{person}{Karoline Wild}, \bibinfo{person}{Uwe Breitenb{\"u}cher}, \bibinfo{person}{Lukas Harzenetter}, \bibinfo{person}{Frank Leymann}, \bibinfo{person}{Daniel Vietz}, {and} \bibinfo{person}{Michael Zimmermann}.} \bibinfo{year}{2020}\natexlab{}.
\newblock \showarticletitle{TOSCA4QC: two modeling styles for TOSCA to automate the deployment and orchestration of quantum applications}. In \bibinfo{booktitle}{\emph{2020 IEEE 24th International Enterprise Distributed Object Computing Conference (EDOC)}}. \bibinfo{publisher}{IEEE}, \bibinfo{address}{Eindhoven, Netherlands}, \bibinfo{pages}{125--134}.
\newblock
\href{https://doi.org/10.1109/EDOC49727.2020.00024}{doi:\nolinkurl{10.1109/EDOC49727.2020.00024}}


\bibitem[Zettili and Zahed(2003)]%
        {Zettili2009}
\bibfield{author}{\bibinfo{person}{Nouredine Zettili} {and} \bibinfo{person}{Ismail Zahed}.} \bibinfo{year}{2003}\natexlab{}.
\newblock \showarticletitle{Quantum Mechanics: Concepts and Applications}.
\newblock \bibinfo{journal}{\emph{American Journal of Physics}} \bibinfo{volume}{71}, \bibinfo{number}{1} (\bibinfo{year}{2003}), \bibinfo{pages}{93--93}.
\newblock
\showISSN{0002-9505}
\href{https://doi.org/10.1119/1.1522702}{doi:\nolinkurl{10.1119/1.1522702}}


\bibitem[Zhang et~al\mbox{.}(2014)]%
        {Zhang2014}
\bibfield{author}{\bibinfo{person}{Yun Zhang}, \bibinfo{person}{Ram Krishnan}, {and} \bibinfo{person}{Ravi Sandhu}.} \bibinfo{year}{2014}\natexlab{}.
\newblock \showarticletitle{Secure Information and Resource Sharing in Cloud Infrastructure as a Service}. In \bibinfo{booktitle}{\emph{Proceedings of the 2014 ACM Workshop on Information Sharing \& Collaborative Security}} (Scottsdale, Arizona, USA) \emph{(\bibinfo{series}{WISCS '14})}. \bibinfo{publisher}{Association for Computing Machinery}, \bibinfo{address}{New York, NY, USA}, \bibinfo{pages}{81--–90}.
\newblock
\showISBNx{9781450331517}
\href{https://doi.org/10.1145/2663876.2663884}{doi:\nolinkurl{10.1145/2663876.2663884}}


\bibitem[Zhao(2020)]%
        {Zhao2020}
\bibfield{author}{\bibinfo{person}{Jianjun Zhao}.} \bibinfo{year}{2020}\natexlab{}.
\newblock \showarticletitle{Quantum software engineering: Landscapes and horizons}.
\newblock \bibinfo{journal}{\emph{arXiv preprint arXiv:2007.07047}} \bibinfo{volume}{1}, \bibinfo{number}{1} (\bibinfo{year}{2020}), \bibinfo{pages}{1--34}.
\newblock
\href{https://doi.org/10.48550/arXiv.2007.07047}{doi:\nolinkurl{10.48550/arXiv.2007.07047}}


\bibitem[Álvaro M.~Aparicio-Morales et~al\mbox{.}(2024)]%
        {AparicioMorales2024}
\bibfield{author}{\bibinfo{person}{Álvaro M.~Aparicio-Morales}, \bibinfo{person}{Enrique Moguel}, \bibinfo{person}{José Garcia-Alonso}, \bibinfo{person}{Alejandro Fernandez}, \bibinfo{person}{Luis~Mariano Bibbo}, {and} \bibinfo{person}{Juan~M. Murillo}.} \bibinfo{year}{2024}\natexlab{}.
\newblock \showarticletitle{Oferta y demanda en la formación de personal para la ingeniería de software cuántico}.
\newblock \bibinfo{journal}{\emph{Memoria Investigaciones en Ingeniería}} \bibinfo{volume}{1}, \bibinfo{number}{1} (\bibinfo{year}{2024}), \bibinfo{pages}{248--256}.
\newblock
Issue 27.
\showISSN{2301-1106}
\href{https://doi.org/10.36561/ING.27.16}{doi:\nolinkurl{10.36561/ING.27.16}}


\end{thebibliography}


\end{document}